\documentclass[12pt]{article}

\usepackage{amsmath,amssymb}

\newcommand{\sca}{\ensuremath{\mathcal{A}}}
\newcommand{\scb}{\ensuremath{\mathcal{B}}}
\newcommand{\scd}{\ensuremath{\mathcal{D}}}
\newcommand{\scg}{\ensuremath{\mathcal{G}}}
\newcommand{\sch}{\ensuremath{\mathcal{H}}}
\newcommand{\scs}{\ensuremath{\mathcal{S}}}
\newcommand{\scx}{\ensuremath{\mathcal{X}}}
\newcommand{\scy}{\ensuremath{\mathcal{Y}}}

\newcommand{\sone}{\ensuremath{\mathcal{S}_{1}}}

\newcommand{\dera}{\ensuremath{Der(\mathcal{A})}}
\newcommand{\sdera}{\ensuremath{SDer(\mathcal{A})}}

\newcommand{\phst}{\ensuremath{\Phi_*}}
\newcommand{\phstup}{\ensuremath{\Phi^*}}

\newcommand{\omcl}{\ensuremath{\omega_{cl}}}

\newcommand{\gstar}{\ensuremath{\mathcal{G}^*}}

\begin{document}
\begin{center} \textbf{\large A stepwise planned approach to the solution
of Hilbert's sixth problem. I : noncommutative symplectic geometry
and Hamiltonian mechanics}

\vspace{.15in} \textbf{Tulsi Dass}

 Indian Statistical Institute, Delhi Centre, 7, SJS
Sansanwal Marg, New Delhi, 110016, India.

 tulsi@isid.ac.in; tulsi@iitk.ac.in

\vspace{.3in} \textbf{Abstract} \end{center}

  This series of  papers is devoted to an open-ended
project aimed at the solution of Hilbert's sixth problem (concerning
joint axiomatization of physics and probability theory) proposed to
be constructed in the framework of an all-embracing mechanics. In
this first paper, the bare skeleton of such a mechanics is
constructed in the form of noncommutative Hamiltonian mechanics (NHM)
which combines elements of noncommutative symplectic geometry and
noncommutative probability in the framework of topological
superalgebras; it includes, besides NHM basics, a  treatment of
Lie group actions in NHM and noncommutative analogues of the
momentum map, Poincar$\acute{e}$-Cartan  form and the symplectic
version of Noether's theorem. Canonically induced symplectic
structure on the (skew) tensor product of two symplectic superalgebras
(needed in the description of interaction between systems) is shown to
exist if and only if either both system superalgebras are
supercommutative or both non-supercommutative with a `quantum
symplectic structure' characterized by a \emph{universal}
Planck type constant; the presence of such a universal constant is,
therefore, \emph{dictated} by the formalism. This provides proper
foundation for an autonomous development of quantum mechanics as a
universal mechanics.

\newpage
 \noindent \textbf{\large Contents}

\vspace{.12in} \noindent \textbf{1. Introduction}

\noindent \textbf{2. Superderivation-based Differential Calculus}
 \begin{description}
 \item[2.1] Superalgebras and superderivations
 \item[2.2] Noncommutative differential forms
 \item[2.3] Induced mappings on differential forms
 \item[2.4] A generalization of the DVNCG scheme
 \item[2.5] Superderivations and differential forms on tensor
            products of superalgebras
 \end{description}
\noindent \textbf{3. Noncommutative Symplectic Geometry and Hamiltonian Mechanics}
\begin{description}
\item[3.1] Symplectic structures; Poisson brackets
\item[3.2] Reality properties of the symplectic form and the Poisson bracket
\item[3.3] Special superalgebras; the canonical symplectic form
\item[3.4] Noncommutative Hamiltonian mechanics
           \begin{description}
           \item[3.4.1] The system algebra and states
           \item[3.4.2] Dynamics
           \item[3.4.3] Eqivalent descriptions; Symmetries and conservation
                        laws
           \item[3.4.4] Classical Hamiltonian mechanics and traditional
                        Hilbert space quantum mechanics as subdisciplines
                        of NHM
           \end{description}
\item[3.5] Symplectic actions of Lie groups in NHM
\item[3.6] The noncommutative momentum map
\item[3.7] Generalized symplectic structures and Hamiltonian systems
\item[3.8] Augmented symplectics including `time'; the noncommutative analogue
           of Poincar$\acute{e}$-Cartan form
\item[3.9] Noncommutative symplectic version of Noether's theorem
          \end{description}
\noindent \textbf{4. Interacting Systems in Noncommutative Hamiltonian mechanics}
\begin{description}
\item[4.1] The symplectic form and Poisson bracket on the (skew) tensor product
           of two symplectic superalgebras
\item[4.2] Dynamics of interacting systems
\end{description}
\noindent \textbf{5. Concluding Remarks}

\vspace{.1in} \noindent  \textbf{Appendix : Study of Implications of the Equations (85) and (86)}

\vspace{.15in} \begin{center} * \ * \ * \ * \ * \end{center}

\noindent \emph{To construct a sensible theory \\
Of all phenomena \\
You must unify \\
Probability and dynamics.}

\vspace{.3in}
 \noindent \textbf{\large 1. Introduction}

\vspace{.12in} In the famous list of 23 problems of David
Hilbert [54],  the key statement of the sixth problem (henceforth called
Hilprob6), presented under the title `Mathematical Treatment of the
Axioms of Physics', reads :

\vspace{.1in} \noindent  ``To treat in the same manner, by means of
axioms, the physical sciences in which mathematics plays an
important part; in the first  rank are the theory of probabilities
and mechanics.''

\vspace{.1in} \noindent We shall adopt the following reformulation
of Hilprob6 which makes clear its close relation to
 `theory of everything' (TOE) :

\vspace{.1in} ``To evolve an axiomatic scheme covering all physics
including the probabilistic framework employed for the treatment of
statistical aspects of physical phenomena.''

\vspace{.1in} Wightman's article [77] is a decent review of the work
relating to the solution of Hilprob6 up to mid-seventies. It covers
Hilbert's own work in this connection and the main developments
relating to the axiomatization of quantum mechanics (QM) and of
quantum field theory (QFT); details and references may be
found in Wightman's paper.

   Developments during the past four decades, relevant to the present
theme, relate to the following approaches : (i) Traditional gauge
models (Weinberg [76]; Donoghue et al [32]), (ii) Superstring theory
(Green, Schwarz and Witten [45]; Polchinski [64]), (iii) Loop
quantum gravity (LQG)(Ashtekar and Lewandowski [6]; Rovelli [65]),
(iv) Noncommutative geometry (NCG)  based formalism for foundations
of geometry (Connes [20,21])
and fundamental interactions (Chamseddine and Connes [17,18]; Connes
and Marcolli [22]), (v) Local quantum physics or algebraic QFT (Haag
[50]; Araki [4]; Horuzhy [55]; Bratteli and Robinson [13]; Borchers
[10]), and (vi) Constructive field theory (Glimm and Jaffe [43];
Baez, Segal and Zhou [7]). Each of these approaches has its
successes and problems. One expects that, in due course of time,
these approaches will converge to an `all-embracing' formalism
facilitating construction of a TOE which, with appropriate
axiomatization, will yield a solution of Hilprob6.

Among the six approaches above, (v) has yet to accommodate concrete
gauge models, whereas (vi) has yet to show concrete progress in four
space-time dimensions. A common feature of the first four is that
they are \emph{quantization} programmes [i.e. one first develops the
formalism in a classical setting (which may involve NCG) and then
quantizes it]. The author considers this a drawback worth taking
seriously and believes that, if one works, instead, in a framework
which adopts a quantum outlook from the beginning, at least some of
the Problems (those with  a capital `P'; for example, that of
providing, in a coherent scheme,  a rationale for the choice of
gauge group in gauge models) will be easier to solve. [To appreciate
this point better, consider the chances of proving PCT and
spin-statistics theorems by someone employing, instead of the
manifestly covariant formalism of quantum field theory, one in which
one only has tricks to \emph{relativize} nonrelativistic equations.]
The present series of papers is aimed at developing such a
complementary approach (to the construction of the `all-embracing'
formalism ) in which commitment to an autonomous quantum theoretic
treatment is one of the top priorities.

Since all physics is essentially mechanics (every branch of physics
deals with the dynamics of one or the other class of systems), the
desired formalism  must be an elaborate scheme of mechanics (with
elements of probability incorporated). The commitment mentioned
above implies that such a scheme of mechanics must incorporate, at
least as a subdiscipline or in some approximation, an
ambiguity-free autonomous development of QM . One expects that,
such a development will, starting with some appealing basics,
connect smoothly to the traditional Hilbert space QM and
facilitate a satisfactory treatment of quantum-classical
correspondence and measurements.

For identifying appropriate ingredients for the desired formalism,
one must go back to physics basics. Physics is concerned with
making/collecting observations, studying correlations between
observations, theorizing
about those correlations and making theory-based conditional
predictions/retrodictions about observations. The desired formalism
must, therefore, incorporate a rich enough description of
observations so as to adequately cover all these aspects. We are
accustomed to describing observations as geometrical facts. The
formalism must, therefore, have an all-embracing underlying
geometry. An appealing choice for the same is NCG (Connes [19];
Dubois-Violette [34]; Gracia-Bondia, Varilly and Figuerra [44];
Madore [60]; Landi [59]) which is known to
cover traditional/commutative differential geometry as a special
case. It has, moreover, excellent consonance with our theme because
noncommutativity is the hallmark of QM. Indeed, the central point
made in Heisenberg's  paper [52] that marked the birth of QM was
that the kinematics underlying QM must be based on a non-commutative
algebra of observables. This idea was developed into a scheme of
mechanics --- called matrix mechanics --- by Born, Jordan, Dirac and
Heisenberg [12,30,11]. The proper geometrical framework for the
construction of the quantum Poisson brackets of matrix mechanics is
provided by non-commutative symplectic structures treated by
Dubois-Violette and coworkers [34,37,35,36]. The NCG scheme employed
in these works (referred to henceforth as DVNCG) is a
straightforward generalization of the scheme of commutative
differential geometry in which the algebra $C^{\infty}(M)$ of smooth
complex-valued functions on a manifold M is replaced by a general
(not necessarily commutative) complex associative *-algebra \sca \
and the Lie algebra $\scx(M)$ of smooth complex vector fields on M by the
Lie algebra Der(\sca) of derivations of \sca.

The key observation underlying the present work is  the fact that
the *-algebras of the type employed in DVNCG also provide a general
framework for an observable-state based treatment of quantum
probability (Meyer [62]). This allows us to adopt the strategy of
combining elements of noncommutative symplectic geometry and
noncommutative probability in an algebraic framework; this promises
to be a reasonably deep level realization of unification of
probability and dynamics (advocated in the poem above) in the true
spirit of Hilprob6.

The scheme based on normed algebras (Jordan, von Neumann and Wigner
[57]; Segal [70,71]; Haag and Kastler [51]; Haag [50]; Araki [4];
Emch [40,41]; Bratteli and Robinson [13]), although it has an
observable - state framework of the type mentioned above, does not
serve our needs because it is not suitable for a sufficiently
general treatment of noncommutative symplectic geometry. Iguri and
Castagnino [56] have analyzed the prospects of a more general class
of topological algebras (nuclear, barreled b*-algebras) as a
mathematical framework for the formulation of quantum principles
prospectively better than that of the normed algebras. These
algebras accommodate unbounded observables at the abstract level.
Following essentially the `footsteps' of Segal [70], they obtain
some results parallel to those in the C*-algebra theory --- an
extremal decomposition theorem for states, a functional
representation theorem for commutative subalgebras of observables
and an extension of the classical GNS theorem. A paper by Dimakis
and M$\ddot{u}$ller-Hoissen [30] is devoted to constructing
consistent differential calculi on the algebra generated by
operators satisfying the canonical commutation relations and
studying invariance properties of the resulting constructions. These
works are essentially  complementary to the present one where the
emphasis is on the development of noncommutative \emph{Hamiltonian
mechanics}.

Development of such a mechanics in the algebraic framework requires
some augmentation and generalization of DVNCG in the following
respects :

\noindent (i) One needs the noncommutative analogues of the
push-forward and pull-back mappings induced by diffeomorphisms
between manifolds on vector fields and differential forms (which
play important roles in classical symplectic geometry).

\noindent (ii) It is desirable to have  a generalization of DVNCG
based on algebraic pairs $(\sca, \scx)$ where \sca \ is a *-algebra
as above and \scx \ is a Lie subalgebra of Der(\sca). (This is
noncommutative analogue of working on a leaf of a foliation; see
section 2.4.) Such a generalization will be needed in the
development of noncommutative symplectic geometry with sufficient
generality and in the algebraic treatment of general quantum
systems.

\noindent (iii) The developments in DVNCG
and noncommutative symplectic structures mentioned above are
purely algebraic. To facilitate analytical work in the development
of the desired mechanics, a sufficiently general and coherent
topological setting must be employed.

The formalism being aimed at is expected to have Wightman's
axiomatic formalism (involving Bose as well as Fermi fields) and
appropriate algebraic versions of modern gauge theories and
superstring theories as special cases. This necessitates
accommodating fermionic objects on an equal footing with the bosonic
ones; to achieve this, we shall employ (topological)
superalgebras as the basic objects.

In the next section, a superalgebraic version of DVNCG is presented
which incorporates the improvement in the definition of
noncommutative differential forms introduced in [35,36] [i.e.
demanding $\omega(...,KX,...) = K \omega(...,X,...)$ where K is in
the center of the algebra; for notation, see section 2.2] and the
augmentation and generalization of DVNCG mentioned above. In section
3, a straightforward development of noncommutative symplectic
geometry and noncommutative Hamiltonian mechanics (NHM) is
presented. It is in the form of an observable-state based algebraic
treatment of mechanics parallel to that of classical statistical
mechanics; the concept of a classical Hamiltonian system is
generalized to that of an NHM Hamiltonian system. Induced mappings
on differential forms mentioned above play an important role in a
systematic treatment of symplectic mappings and invariance
principles and in defining  equivalence of NHM Hamiltonian systems.
This section also includes a treatment, in the NHM framework, of
symplectic actions of Lie groups and noncommutative analogues
of the momentum map, Poincare-Cartan form and the symplectic version
of Noether's theorem [Theorem(1)].

Section 4 contains the treatment of two interacting systems in the
framework of NHM. An important result obtained there [Theorem (2)]
is that, given two systems represented by symplectic superalgebras
$\sca^{(i)}$ (i = 1,2) with symplectic forms $\omega^{(i)}$, the
`canonically induced' two-form $\omega$ [given by Eq.(78) below] on
the (skew) tensor product $ \sca^{(1)} \otimes \sca^{(2)}$ (the
system algebra of the coupled system) represents a genuine
symplectic structure if and only if either both the superalgebras
are supercommutative or both non-supercommutative with a `quantum
symplectic structure' [i.e. one which gives a Poisson bracket which
is a (super-)commutator up to multiplication by a constant $(i
h_0^{-1})$ where $h_0$ is a real-valued constant of the dimension of
action] characterized by a \emph{universal} parameter $h_0$. It
follows that the formalism, firstly, prohibits a `quantum-classical
interaction', and, secondly, has a natural place for the Planck
constant as a universal constant --- just as special relativity has
a natural place for a universal speed. In fact, the situation here
is substantially better because whereas, in special relativity, the
existence of a universal speed is \emph{postulated}, in NHM, the
existence of a universal Planck-like constant is
\emph{dictated/predicted} by the formalism.

An important result obtained in this treatment is the formula (99)
for the Poisson bracket in the (skew) tensor product of two
symplectic superalgebras.

The last section contains some concluding remarks.

The formalism of NHM needs to be augmented to make a satisfactory
autonomous treatment of quantum systems possible; this augmentation
and an autonomous development of QM in the framework of augmented
NHM will be presented in the next paper in the series.

\noindent \emph{Note.} (i) The work presented in the first two papers
in this series substantially supersedes that in the relevant parts of
earlier (unpublished) papers in the arXiv [24-27].

\noindent (ii) A very brief account of the main developments in the
first two papers in this series was presented in a short lecture in
the ICM2010 meeting [28].

\vspace{.15in} \noindent \textbf{\large 2. Superderivaion-based
Differential Calculus}

In this section, we present essential developments in the
noncommutative differential calculus to be employed in this series
of papers. This is a superalgebraic version of DVNCG augmented and
generalized as mentioned above. Topics covered include some not so
well known results about superalgebras and superderivations,
noncommutative differential forms and some of their transformation
theory and superderivations and differential forms on (skew) tensor
products of superalgebras.

Some good references for the background material for this section
are (apart from the references for DVNCG given above; see also the
review article by Djemai [31]) Greub [46,47]; Pittner [63];
Giachetta, Mangiarotti and Sardanshvily [42]; Scheunert [67,68];
Scheunert and Zhang [69].

\vspace{.1in}  \noindent \emph{Note}. In most applications, the
non-super version of the formalism developed below is adequate; this
can be obtained by simply putting, in the formulas below, all the
epsilons representing parities equal to zero.

\vspace{.12in} \noindent \textbf{2.1 Superalgebras and
superderivations}

We first recall a few basic concepts in  superalgebra. A
\emph{supervector space} is a (complex) vector space $V = V^{(o)}
\oplus V^{(1)} $; a vector $v \in V$ can be uniquely expressed as a
sum $v = v_0 + v_1$ of even and odd vectors; they are assigned
parities $\epsilon(v_0) = 0$ and $\epsilon(v_1) = 1$. Vectors with
definite parity are called homogeneous. We shall denote the parity
of a homogeneous vector $w$ by $\epsilon(w)$ or $\epsilon_w$
according to convenience. A \emph{superalgebra} \sca \ is a
supervector space which is an associative algebra with identity
(denoted as $I_{\sca}$ or simply I); it becomes a
\emph{*-superalgebra} if an antilinear
*-operation or involution $*:\sca \rightarrow \sca$ \ is defined
such that $ (A^*)^* = A$, $(AB)^* = B^*A^*$ for all $A,B \in  \sca,
\ I^* = I$ and, for homogeneous $A \in \sca, \epsilon (A^*) =
\epsilon (A)$. An element $A \in \sca$ \  is called \emph{hermitian}
if $ A^* = A.$

\noindent \emph{Note.} In a couple of earlier versions of this paper
(arXiv : 0909.4606 v1, v2), the following convention  was adopted :
For homogeneous A and B,  $(AB)^* =
\eta_{AB}B^*A^* $ where $\eta_{AB} = (-1)^{\epsilon_A \epsilon_B}.$
This convention, however, does not suit our needs. For example,
given two fermionic annihilation operators a, b, we have, in
traditional quantum field theory, $(ab)^* = b^* a^*$ and not $(ab)^*
= - b^* a^*$.

The \emph{supercommutator} of two homogeneous elements A,B of \sca \
is defined as $[A,B] = AB -\eta_{AB}BA;$ the definition is extended
to general elements by bilinearity. We shall also employ the
notations $[A,B]_{\mp} = AB \mp BA.$ A superalgebra \sca \ is said
to be \emph{supercommutative} if the supercommutator of every pair
of its elements vanishes. The \emph{graded center} of \sca, denoted
as $Z(\sca)$, consists of those elements of \sca \ which have
vanishing supercommutators with all elements of \sca; it is clearly
a supercommutative superalgebra. A \emph{Lie superalgebra} is a
supervector space $\mathcal{L}$ with a \emph{superbracket} operation
$[ \ , \ ]_{\bullet} : \mathcal{L} \times \mathcal{L} \rightarrow
\mathcal{L}$ which is (i) bilinear, (ii) graded skew-symmetric which
means that, for any two homogeneous elements $a,b \in \mathcal{L},
[a,b]_{\bullet} = -\eta_{ab}[b,a]_{\bullet}$ and (iii) satisfies the
\emph{super Jacobi identity}
\begin{eqnarray*}
[a,[b,c]_{\bullet} ]_{\bullet} = [[a,b]_{\bullet} ,c ]_{\bullet} +
\eta_{ab}[b,[a,c]_{\bullet} ]_{\bullet} \
\textnormal{for all} \ a,b,c \in \mathcal{L}.
\end{eqnarray*}
[The subscript  bullet ($\bullet$) has been temporarily introduced
in the notation for a Lie bracket to keep
this object distinct from the supercommutator. We shall generally
omit this subscript; it should be clear from the context whether a
particular square bracket stands for a supercommutator or a Lie
superbracket.]

For analytical work we  need to operate in a topological setting;
in particular, the algebras we employ must be topological algebras
[33,53]. The class of
these generally employed will consist of complete,
Hausdorff, separable locally convex algebras with a jointly
continuous product (with a continuous star operation and
$Z_2$-grading); motivations for this choice are described in
section 3.4. Some instructive material relating to the so-called
`smooth *-algebras' (which are closely related to the *-algebras
we need) may be found in [39].

For algebro-topological matters, we shall generally keep close to
Helemskii [53]; for
some aspects of group actions on locally convex spaces, we
shall take guidance from Yosida [79]. The defining system of
seminorms $\Gamma = \{ p_{\nu}; \nu \in \Lambda \}$  for a
locally convex superalgebra  \sca \ will be taken to be
saturated; it means that, for any finite subset F of $\Lambda$,
the seminorm $p_F$ defined by
\[ p_F (A) = max_{\nu \in F} \ p_{\nu} (A), \ \ A \in \sca \]
is in $\Gamma$. The seminorms $p_{\nu}$ will be taken to be
*-preserving [ i.e. $ p_{\nu} (A^*) = p_{\nu} (A)  \ \forall
A \in \sca $]. For every $ \nu \in \Lambda$, there always
exists [53] a $\mu \in \Lambda$ such that $p_{\nu} (AB) \leq
p_{\mu}(A) p_{\mu}(B)$ for all $A,B \in \sca.$ Mappings
involving topological spaces should be understood as
continuous (which, in the case of multilinear mappings,
shall mean jointly continuous) unless stated otherwise.

A (topological) \emph{(*-)homomorphism} of a superalgebra
\sca \ into \scb \ is a (continuous)
linear mapping $ \Phi : \sca \rightarrow \scb $ which preserves
products, identity elements, parities (and involutions) :
\begin{eqnarray*}
\Phi (AB) = \Phi (A) \Phi (B), \hspace{.12in} \Phi (I_{\sca}) =
I_{\scb}, \hspace{.12in} \epsilon (\Phi (A)) = \epsilon (A),
\hspace{.12in} \Phi (A^*) = (\Phi (A))^*;
\end{eqnarray*}
if it is, moreover, bijective, it is called a (topological)
\emph{(*-)isomorphism}.

A (homogeneous) \emph{superderivation} of a superalgebra \sca \ is a
linear map $X: \sca \rightarrow \sca$ such that $X(AB) = X(A)B +
\eta_{XA} AX(B)$ for all homogeneous A and all B in \sca. The
\emph{multiplication operator} $\mu$ on \sca \ associates, with
every $A \in \sca$, a linear mapping $\mu(A) : \sca \rightarrow
\sca$ such that  $\mu (A)B = AB$ for all $B \in \sca.$

\vspace{.1in} \noindent \textbf{Proposition 2.1.} \emph{Given a
superalgebra \sca, a necessary and sufficient condition that a
linear map $X: \sca \rightarrow \sca$ is a homogeneous
superderivation is}
\begin{eqnarray}
X \circ \mu(A) - \eta_{XA}\  \mu(A) \circ X = \mu(X(A)) \
\textnormal{for all homogeneous} \  A \in \sca.
\end{eqnarray}

\noindent \emph{Proof.} Eq.(1) gives, on making each side act on a
general element B of \sca, precisely the equation defining a
homogeneous superderivation X above. $\Box$

\vspace{.1in} \noindent The space SDer(\sca) [$ = \sdera^{(0)}
\oplus \sdera^{(1)}$] of all superderivations of \sca \ constitutes
a Lie superalgebra with $[ X, Y ]_{\bullet} = [X, Y]$. This space
will be understood to have the subspace topology of the space
L(\sca) of continuous linear mappings  of \sca \
into itself equipped with the topology of bounded convergence
(Yosida [79]; p. 110). The \emph{inner
superderivations} $D_A$ defined by $D_AB = [A,B]$ satisfy the
relation $[D_A,D_B] = D_{[A,B]}$ and constitute a Lie
sub-superalgebra ISDer(\sca) of SDer(\sca).

The following two propositions (easily proved) list superalgebraic
generalizations of some formulas appearing in Ref.[35].

\vspace{.1in} \noindent \textbf{Proposition 2.2.} \emph{Given a
superalgebra \sca,  $K \in Z(\sca)$ and $X,Y \in  SDer(\sca)$, we
have (i) $X(K) \in Z(\sca) $, (ii) $ KX \in SDer(\sca)$ [with
$(KX)(A) \equiv K [X(A)]$ for all $A \in \sca$], and (iii) with X,K
homogeneous, the relation}
\begin{eqnarray}
[X,KY] = X(K)Y + \eta_{XK}K[X,Y].
\end{eqnarray}

\noindent It follows as an immediate corollary that \sdera \ is a
Z(\sca)-module.

\vspace{.1in}  An involution * on \sdera \ is defined by the
relation $ X^*(A) = [X(A^*)]^*.$

\vspace{.1in} \noindent \textbf{Proposition 2.3.} \emph{In obvious
notation}

\vspace{.1in} \noindent (i) $[X,Y]^* = [X^*,Y^*]; \ \ (ii)\ (D_A)^*
= -D_{A^*}.$

\vspace{.1in}  A (topological) superalgebra-isomorphism
$\Phi : \sca \rightarrow
\scb$ induces a (continuous) linear mapping
\begin{eqnarray} \phst : \sdera \rightarrow SDer(\scb)
 \ \ \textnormal{given by} \ \  (\phst X)(B) = \Phi(X[\Phi^{-1}(B)])
\end{eqnarray}
for all $X\in \sdera $ and $B \in \scb.$ It is the analogue (and a
generalization) of the push-forward mapping induced by a
diffeomorphism between two manifolds on the vector fields.

\vspace{.1in} \noindent \textbf{Proposition 2.4.} \emph{With $\Phi$
as above and $\Psi : \scb \rightarrow \mathcal{C}$ we have }
\begin{eqnarray}
& (i) & \ (\Psi \circ \Phi)_* = \Psi_* \circ \Phi_*; \hspace{.12in}
(ii)
\ \Phi_* [X,Y] = [\phst X, \phst Y]; \nonumber \\
& (iii)&  [\phst(X)]^* = \phst(X^*).
\end{eqnarray}
\emph{Proof.} (i) For any $X \in Sder(\sca)$ and  $C \in
\mathcal{C}$,
\begin{eqnarray*}
[(\Psi \circ \Phi)_*X](C) & = & (\Psi \circ \Phi)(X[(\Psi \circ
\Phi)^{-1}(C)]) = \Psi [\Phi (X [\Phi^{-1}(\Psi^{-1}(C))])] \\
                          & = & \Psi [ (\Phi_*X)(\Psi^{-1}(C)]
                           =  [\Psi_*(\Phi_*X)](C).
\end{eqnarray*}
(ii) It is clearly adequate to prove it for homogeneous $X,Y \in
\sdera$. For any $B \in \mathcal{B}$, we have
\begin{eqnarray*}
( \phst[X,Y])(B) & = & \Phi ([X,Y](\Phi^{-1}(B))) \\
                 & = & \Phi[X (Y [\Phi^{-1}(B)])] - \eta_{XY} \Phi [Y(X[\Phi^{-1}(B)])].
\end{eqnarray*}
Now, inserting  $\Phi^{-1} \circ \Phi$ between X and Y in each of
the two terms on the right, the right hand side is easily seen to be
\[ (\Phi_* X)[(\Phi_* Y)(B)] - \eta_{XY} (\Phi_* Y) [(\Phi_* X)(B)].
 \]

\vspace{.1in} \noindent (iii) For any  $ B \in \scb, $ we have
\begin{eqnarray*} [\phst(X)]^*(B) & = & [\phst(X)(B^*)]^*
=  [\Phi(X[\Phi^{-1}(B^*)])]^* \\
& = & \Phi [(X[\Phi^{-1}(B^*)])^*] =  \Phi (X^*[\Phi^{-1}(B)])  =
(\phst X^*)(B) \ \ \Box \end{eqnarray*}

\vspace{.1in} \noindent  \textbf{Corollary 2.5.} \emph{The mapping
\phst \ defined by Eq.(3) is an involution preserving Lie
superalgebra isomorphism.}

\vspace{.12in} \noindent \textbf{2.2 Noncommutative differential
forms}

Main developments in this subsection are parallel to those in
(Grosse and Reiter [48]) who have generalized the treatment of
differential geometry of matrix algebras in [37] to supermatrix
algebras. Some related work on supermatrix geometry has also
appeared in [38,58].

We shall construct a differential calculus on a superalgebra \sca \
by employing (the super-version of) a subclass of
Chevalley-Eilenberg cochains [16,75,42].
It is, in fact, somewhat simpler to introduce this subclass
directly; however, since we shall need a couple of results
relating to another subclass later, it appears worthwhile spending
an extra para on a brief treatment of the general case.

Let $\mathcal{G}$ be a Lie algebra over the field $\mathbb{K}$
[which may be $\mathbb{R}$ or $\mathbb{C}$] and V a
$\mathcal{G}$-module which means it is a  vector space over
$\mathbb{K}$ having defined on it a $\mathcal{G}$-action associating
a linear mapping $\Psi(\xi)$ on V with every element $\xi$ of
$\mathcal{G}$ such that
\[ \Psi(0) = id_V \ \ \textnormal{and} \ \ \Psi([\xi,\eta]) =
\Psi(\xi) \circ \Psi(\eta) - \Psi(\eta) \circ \Psi(\xi) \
\textnormal{for all} \ \xi, \eta \in \mathcal{G} \] where $id_V$ is
the identity mapping on V. A V-valued p-cochain $\lambda^{(p)}$ of
$\mathcal{G}$ (p = 1,2,...) is a skew-symmetric multilinear map from
$\mathcal{G}^p$ into V. These cochains constitute a vector space
$C^p(\mathcal{G},V).$  One defines $C^0(\scg,V) = V$. The coboundary
operator $d : C^p(\mathcal{G},V) \rightarrow C^{p+1}(\scg,V)$
defined by
\begin{eqnarray} (d\lambda^{(p)})(\xi_0,\xi_1,..,\xi_p)  =
\sum_{i=0}^{p} (-1)^i
\Psi(\xi_i)[\lambda^{(p)}(\xi_0,..,\hat{\xi}_i,.., \xi_p)] +
\nonumber
\\
\sum_{0 \leq i<j \leq p}(-1)^j \lambda^{(p)}(\xi_0,..,\xi_{i-1},
[\xi_i,\xi_j],\xi_{i+1},..,\hat{\xi}_j,..,\xi_p) \end{eqnarray} (the
hat indicates omission) for
 $\xi_0,..,\xi_p \in \mathcal{G}$ satisfies the condition $d^2 = 0$.
Defining \[C(\scg,V) = \bigoplus_{p\geq 0} C^p(\scg,V),\] the pair
($C(\scg,V), d)$ constitutes a cochain complex. The subspaces of
$C^p(\mathcal{G},V)$ consisting of closed cochains (cocycles)[i.e.
those $\lambda^{p}$ satisfying $d\lambda^{p} =0$] and exact cochains
(coboundaries) [i.e. those $\lambda^p$ satisfying $\lambda^p = d
\mu^{p-1}$ for some (p-1)-cochain $\mu^{p-1}$] are denoted as
$Z^p(\mathcal{G},V)$ and $B^p(\scg,V)$ respectively; the quotient
space $H^p(\scg,V) \equiv Z^p(\scg,V)/B^p(\scg,V)$ is called the
p-th cohomology space of \scg \ with coefficients in V.

For the special case of the trivial action of \scg \ on V [i.e.
$\Psi(\xi) = 0 \ \forall \xi \in \scg$], a subscript zero is
attached to these spaces [$C^p_0(\scg,V)$ etc].  In this case, we
record, for future use, the form Eq.(5) takes for p = 1,2 :
\begin{eqnarray} (d \lambda^{(1)})(\xi_0,\xi_1) & = & -
\lambda^{(1)}([\xi_0,\xi_1]) \nonumber \\
(d \lambda^{(2)}) (\xi_0,\xi_1,\xi_2)  & = & -
[\lambda^{(2)}([\xi_0,\xi_1], \xi_2) + \textnormal{cyclic terms in}
\ \xi_0, \xi_1, \xi_2]. \end{eqnarray}

Recalling that the classical differential p-forms on a manifold M
are defined as skew-symmetric multilinear maps of $\mathcal{X}(M)^p
$ into $C^{\infty}(M)$, the De Rham complex of classical
differential forms can be seen as a special case of
Chevalley-Eilenberg complex with $\scg = \scx(M)$, V =
$C^{\infty}(M)$ and $\Psi (X) (f) = X(f)$ in obvious notation. [For
the relevant algebraic definition of the classical exterior
derivative, see Matsushima [61], p. 140; it is Eq.(5) above with
$\lambda^{(p)}$ a traditional differential p-form and $\xi_j$s
vector fields.] Replacing $C^{\infty}(M)$ by a superalgebra
[complex, associative, unital (i.e. possessing a unit element), not
necessarily supercommutative] \sca \ and $\scx(M)$ by \sdera \ [so
that $\Psi(X)(A) = X(A)$ for all $X \in \sdera$ and $A \in \sca$], a
natural choice for the space of noncommutative differential p-forms
is the space of \sca-valued p-cochains of \sdera :
\begin{eqnarray*}  C^p(SDer\sca,\sca)\  [= C^{p,0}(\sdera,\sca) \oplus
C^{p,1}(\sdera,\sca)] \end{eqnarray*}  with $C^0 (\sdera, \sca) =
\sca$. For $\omega \in C^{p,s}(\sdera, \sca),$  and homogeneous $X,Y
\in \sdera$, we have
\begin{eqnarray}
\omega(..,X,Y,..) = - \eta_{XY} \omega(..,Y,X,..).
\end{eqnarray}
For a general permutation $\sigma$ of the arguments of $\omega$, we
have
\begin{eqnarray*}
\omega(X_{\sigma(1)},..,X_{\sigma(p)}) = \kappa_{\sigma}
\gamma_p(\sigma; \epsilon_{X_1},..,\epsilon_{X_p})
\omega(X_1,..,X_p)
\end{eqnarray*}
where $\kappa_{\sigma}$ is the parity of the permutation $\sigma$
and
\begin{eqnarray*}
\gamma_p(\sigma; s_1,..,s_p) = \prod_{\begin{array}{c}
j,k= 1,..,p; \\
j<k,\sigma^{-1}(j) > \sigma^{-1}(k) \end{array}} (-1)^{s_j s_k}.
\end{eqnarray*}
The space $C^p(SDer\sca,\sca)$ is a left $\sca-$module with the obvious left
\sca-action given by $(A \alpha)(X_1,..,X_p) = A [\alpha (X_1,..,X_p)].$

The \emph{exterior product}
\begin{eqnarray*} \wedge : C^{p,r}(\sdera,\sca) \times C^{q,s}(\sdera,\sca)
\rightarrow C^{p+q,r+s}(\sdera,\sca) \end{eqnarray*} is defined as
\begin{eqnarray*}
(\alpha \wedge \beta)(X_1,..,X_{p+q}) = \frac{1}{p!q!}\sum_{\sigma
\in \mathcal{S}_{p+q}} \kappa_{\sigma} \gamma_{p+q}(\sigma;
\epsilon_{X_1},..,\epsilon_{X_{p+q}})
(-1)^{s \sum_{j=1}^p \epsilon_{X_{\sigma(j)}}}  \nonumber \\
\alpha(X_{\sigma(1)},..,X_{\sigma(p)}) \beta(X_{\sigma(p+1)},..,
X_{\sigma(p+q)}).
\end{eqnarray*}
[When $ \alpha = A,$ a 0-form, it is easily seen that
$A \wedge \beta = A \beta.$]
With this product, the graded vector space
\begin{eqnarray*}
C(\sdera,\sca) = \bigoplus_{p\geq0}C^p(\sdera,\sca)
\end{eqnarray*}
 becomes an $ \mathbb{N}_0 \times Z_2$-bigraded complex algebra.
(Here $ \mathbb{N}_0 $ is the set of non-negative integers.)

An involution $* : C^{p,r}(\sdera,\sca) \rightarrow
C^{p,r}(\sdera,\sca) $ is defined by the relation
$\omega^*(X_1,..,X_p) = [\omega(X_1^*,..,X_p^*)]^*;$
 $\omega$ is said to be real (imaginary) if $ \omega^* =
 \omega (- \omega)$.

The Lie superalgebra \sdera \  acts on itself and on C(\sdera,\sca)
through \emph{Lie derivatives}. For each $Y \in \sdera^{(r)}$, one
defines linear mappings $L_Y :\sdera^{(s)} \rightarrow
\sdera^{(r+s)}$ and $L_Y : C^{p,s}(\sdera, \sca)\rightarrow
C^{p,r+s}(\sdera,\sca) $ \ such that the following three conditions
hold (with obvious notation): \begin{eqnarray*} L_Y(A)  =  Y(A); \
 L_Y[X(A)]  =  (L_YX)(A) + \eta_{XY}X[L_Y(A)]; \\
L_Y [\omega(X_1,..,X_p)]   = (L_Y\omega)(X_1,..,X_p) +
\sum_{i=1}^p (-1)^{\epsilon_Y(\epsilon_{\omega}
                           + \epsilon_{X_1} + ..
+\epsilon_{X_{i-1}})}. \\
.\omega(X_1,..,X_{i-1},L_YX_i,X_{i+1},..,X_p).
\end{eqnarray*}
The first two conditions give
\begin{eqnarray*}
L_Y X = [Y,X]
\end{eqnarray*}
which, along with the third, gives
\begin{eqnarray}
(L_Y \omega)(X_1,..,X_p)  =  Y[\omega(X_1,..,X_p)]
                          - \sum_{i=1}^p (-1)^{\epsilon_Y(\epsilon_{\omega} + \epsilon_{X_1} + .. +
\epsilon_{X_{i-1}})}. \nonumber \\
.\omega (X_1,..,X_{i-1}, [Y,X_i],X_{i+1},..,X_p).
\end{eqnarray}
Two important properties of the Lie derivative are, in obvious
notation,
\begin{eqnarray*}
[L_X,L_Y] & = & L_{[X,Y]} \\
L_Y(\alpha \wedge \beta) & = & (L_Y \alpha) \wedge \beta + \eta_{Y
\alpha} \ \alpha \wedge (L_Y \beta).
\end{eqnarray*}

For each $X \in \sdera^{(r)} $, we define the \emph{interior
product} \\
$ i_X : C^{p,s}(\sdera,\sca) \rightarrow C^{p-1,r+s}(\sdera,\sca) $
 ( for $p \geq 1$) by
\begin{eqnarray*}
(i_X \omega)(X_1,..,X_{p-1}) = \omega (X,X_1,..,X_{p-1}).
\end{eqnarray*}
One defines $i_X(A) = 0 $ for all $A \in \sca \equiv
C^0(\sdera,\sca).$ Some important properties of the interior product
are : \[ i_X \circ i_Y + \eta_{XY} i_Y \circ i_X = 0; \]
\begin{eqnarray}
(L_Y \circ i_X - i_X \circ L_Y) \omega = \eta_{Y \omega} i_{[Y,X]}
\omega. \ \ \ \
\end{eqnarray}

The \emph{exterior derivative} $d : C(SDer(\sca), \sca) \rightarrow
C(SDer(\sca), \sca)$ such that $ d C^{p,r}(\sdera,\sca) \subset
C^{p+1,r}(\sdera,\sca) $ is defined as follows :
\begin{eqnarray}
(d \omega)(X_0,X_1,..,X_p) = \sum_{i=0}^{p} (-1)^{i+ a_i} X_i [
\omega(X_0,..,\hat{X}_{i},..,X_p)]
\nonumber \\
+ \sum_{0 \leq i < j \leq p}(-1)^{j+ b_{ij}} \omega (X_0,..,
X_{i-1}, [X_i,X_j],X_{i+1}, ..,\hat{X}_j,..,X_p)
\end{eqnarray}
where
\begin{eqnarray*}
a_i = \epsilon_{X_i} (\epsilon_{\omega} + \sum_{k=0}^{i-1}
\epsilon_{X_k}); \ b_{ij} = \epsilon_{X_j}\sum_{k=i+1}^{j-1}
\epsilon_{X_k}. \end{eqnarray*} [Eq.(10) is clearly a special case
of (the super-version of) Eq.(5).] It satisfies the relations
(i) $d^2 (= d \circ d) = 0 $,
(ii) $d (\omega^*) = (d \omega)^*$, (iii) $d \circ L_Y = L_Y \circ
d, $ and (iv) the relation
\begin{eqnarray*}
 d (\alpha \wedge \beta) = (d \alpha) \wedge \beta + (-1)^p \alpha
\wedge (d \beta) \hspace{1.7in} (10+)
\end{eqnarray*}
where $\alpha$ is a p-cochain. The first of these equations  shows
that the pair (C(\sdera, \sca), d) constitutes a cochain complex. We
also have the important relation
\begin{eqnarray}
(i_X \circ d + d \circ i_X ) \omega = \eta_{X \omega} \ L_X \omega.
\end{eqnarray}

In Eq.(10+), when $\alpha = A,$ a 0-form, we have
\begin{eqnarray*}
d(A \beta) = (dA) \wedge \beta + A \ (d \beta);  \hspace{2.1in} (11+)
\end{eqnarray*}
if, moreover, $\beta = d \gamma,$ we have
\begin{eqnarray*}
d (A d \gamma) = (dA) \wedge d \gamma. \hspace{2.3in} (11++)
\end{eqnarray*}

Taking clue from [35,36] (where the subcomplex of $Z(\sca)$-linear
cochains [Z(\sca) being, in the notation of these papers, the center
of the algebra \sca] was adopted as the space of differential
forms), we consider the subset $\Omega(\sca)$ of $C(\sdera, \sca)$
consisting of $Z(\sca)$-linear cochains [where Z(\sca) is now the
graded center of the superalgebra \sca]. Eq.(2) ensures that this
subset is closed under the action of d and, therefore, a subcomplex.
We shall take this space to be the space of differential forms in
subsequent geometrical work. We have, of course,
\begin{eqnarray*}
\Omega(\sca) = \bigoplus_{p \geq 0}\Omega^p(\sca)
\end{eqnarray*}
with $\Omega^0(\sca) = \sca$ and $\Omega^p(\sca) =
\Omega^{p,0}(\sca) \oplus \Omega^{p,1}(\sca)$ for all $ p \geq 0.$

\vspace{.12in} \noindent \textbf{2.3 Induced mappings on differential
forms}

\vspace{.12in} A (topological) superalgebra *-isomorphism $ \Phi : \sca
\rightarrow \scb $ induces, besides the Lie superalgebra-isomorphism
$ \phst :\sdera \rightarrow SDer(\scb),$ a (continuous) linear
mapping
\[ \phstup : C^{p,s}(SDer (\scb), \scb) \rightarrow C^{p,s}(\sdera,\sca)\]
given, for all $\omega \in C^{p,s}(SDer(\scb), \scb)$ and $ X_1, ..,
X_p \in \sdera,$  by
\begin{eqnarray}
(\phstup \omega)(X_1,..,X_p) = \Phi^{-1} [ \omega ( \phst X_1,..,
\phst X_p)].
\end{eqnarray}
The mapping $ \Phi$ preserves (bijectively) all the algebraic
relations. Eq.(3) shows that \phst \ preserves $ Z(\sca)$-linear
combinations of the superderivations. It follows that \phstup \ maps
differential forms (with their restricted definition given at the
end of section 2.2) onto differential forms. These mappings are
analogues (and generalizations) of the pull-back mappings on
differential forms (on manifolds) induced by diffeomorphisms.

\vspace{.1in} \noindent \textbf{Proposition 2.6.} \emph{With $\Phi$
and $\Psi$ as in proposition 2.4, $\alpha \in C^{p,r}(SDer(\scb),
\scb)$ and $ \beta \in C^{q,s}(SDer(\scb),\scb),$ we have}
\begin{eqnarray}
(i) \ (\Psi \circ \Phi)^* = \Phi^* \circ \Psi^*; \ \ (ii) \ \phstup
(\alpha \wedge \beta) = (\phstup \alpha) \wedge (\phstup \beta);
\end{eqnarray}
\begin{eqnarray}
(iii) \ \phstup (d \alpha ) = d ( \phstup \alpha );
\end{eqnarray}
\begin{eqnarray} (iv) \  (\phstup \omega)^* = \phstup (\omega^*).
\end{eqnarray}
[Reader, please note the two different uses of * in Eq.(15).]

\vspace{.1in} \noindent \emph{Proof.} (i) For $ \omega \in
C^{p,s}(SDer(\mathcal{C}), \mathcal{C})$ and $ X_1,..,X_p \in
SDer(\sca)$,
\begin{eqnarray*}
[(\Psi \circ \Phi)^* \omega] (X_1,..,X_p) & = & (\Phi^{-1} \circ
\Psi^{-1})[\omega(\Psi_*(\Phi_*X_1),..,
       \Psi_*(\Phi_*X_p))] \\
& = & \Phi^{-1} [(\Psi^* \omega)(\Phi_*X_1,..,\Phi_*X_p)] \\
& = & [\Phi^*(\Psi^* \omega)] (X_1,..,X_p).
\end{eqnarray*}

\noindent (ii)  For $ X_1,..,X_{p+q} \in \sdera$, we have
\begin{eqnarray*}
[\Phi^*(\alpha \wedge \beta)](X_1,.., X_{p+q}) & = & \Phi^{-1}[
(\alpha \wedge \beta) (\phst X_1,..,\phst X_{p+q})].
\end{eqnarray*}
Expanding the wedge product and noting that \[ \Phi^{-1}[\alpha
(..)\beta(..)] = \Phi^{-1}[\alpha(..)]. \Phi^{-1}[\beta(..)],
\] the right hand side is easily seen to be equal to $ [(\phstup
\alpha)\wedge (\phstup \beta)](X_1,..,X_{p+q}).$

 \noindent \vspace{.1in} (iii) We have
\begin{eqnarray*}
[\phstup (d \alpha)](X_0,..,X_p) = \Phi^{-1} [(d \alpha)(\phst
X_0,..,\phst X_p)].
\end{eqnarray*}
Using Eq.(10) for $d \alpha$ and noting that, in the first sum,
\begin{eqnarray*}
\Phi^{-1}[(\phst X_i)(\alpha (\phst X_0,..))
& = & \Phi^{-1} [\Phi (X_i [\Phi^{-1}(\alpha(\phst X_0,..))]]  \\
& = & X_i [(\phstup \alpha) (X_0,..)]
\end{eqnarray*}
and (recalling that $[\Phi_*X_i, \Phi_*X_j] = \Phi_*[X_i, X_j]$), in
the second sum,
\begin{eqnarray*}
\Phi^{-1}[ \alpha (\Phi_*X_0,..,\Phi_*X_i, \Phi_*[X_i,
X_j],..,\Phi_*\hat{X_j},..,\Phi_*X_p)  \\
(\Phi^* \alpha)(X_0,..,X_i, [X_i, X_j],..,\hat{X}_j,..,X_p),
\end{eqnarray*}
it is easily seen that the left hand side of Eq.(14), evaluated at
$(X_0,..,X_p)$, equals $[(d (\phstup \alpha)](X_0,..,X_p).$

\vspace{.1in} \noindent (iv) With the notation in the proof of (i)
above, we have
\begin{eqnarray*} (\phstup \omega)^*(X_1,...,X_p)
& = & [(\phstup \omega)(X_1^*,...,X_p^*)]^*   \\
& = & ( \Phi^{-1}[\omega(\phst X_1^*,...,\phst X_p^*)])^* \\
& = & \Phi^{-1} [\omega ((\phst X_1)^*,...,(\phst X_p)^*]^* \\
& = & \Phi^{-1} [\omega^*(\phst X_1,...,\phst X_p)] \\
& = & (\phstup \omega^*) (X_1,...,X_p).  \ \ \Box  \end{eqnarray*}

\vspace{.1in} Taking \sca \ to be a topological algebra as mentioned
above, let $ \Phi_t : \sca \rightarrow \sca $ ($t \in I_0$ where
$I_0$ is an interval in $\mathbb{R}$ containing the origin) be a
one-parameter family  of transformations (i.e. topological superalgebra
isomorphisms) satisfying the conditions
\begin{eqnarray*} \Phi_s \circ \Phi_t = \Phi_{s+t} \ \textnormal{for
all} \ s, t, s+t \in I_0; \ \Phi_0 = id_{\sca}. \end{eqnarray*} This
family is (i) a group if $I_0 = \mathbb{R}$, (ii) a semigroup if
$I_0 = [0, \infty)$ and (iii) a local group if $I_0$ is an open
interval containing the origin. (In the first two cases, the
condition $ s + t \in I_0$ above is obviously redundant.) We shall
restrict this family to a subclass such as to permit construction
of an infinitesimal generator for it. To this end, we follow Yosida
and impose the following two conditions on the family
$ \{ \Phi_t \}$:

\noindent (i) the $(C_0)$ condition :
\[ lim_{t \rightarrow t_0} \Phi_t(A) = \Phi_{t_0}(A) \
\textnormal{for all} \ t_0 \in I_0 \ \textnormal{and
all} \ A \in \sca; \]

\noindent (ii) the condition of equicontinuity in t which means
that, for any continuous seminorm p on \sca, there exists a
continuous seminorm q on \sca \ such that
\[p (\Phi_t(A)) \leq q(A) \ \textnormal{for all} \ t \in I_0
\ \textnormal{and all} \ A \in \sca. \]
(For a general treatment of equicontinuous
mappings, see Treves [74].)

\noindent \emph{Note.} A treatment parallel to Yosida requires,
in addition, sequential completeness for \sca \ (which is a
weaker condition than completeness which we have already assumed).

\noindent One can now define a linear, even mapping $g : D(g)
\rightarrow \sca $ where the
domain D(g) of g is a dense subset of \sca \  such that, for
all $A \in D(g)$, the limit
\[ g(A) \equiv lim_{t \rightarrow 0} \  t^{-1} [\Phi_t(A) - A] \]
exists. For small t, we have $\Phi_t (A) \simeq A + t g(A).$ The
condition $ \Phi_t (AB) = \Phi_t (A) \Phi_t (B) $ gives $ g(AB)
= g(A) B + A g(B)$ implying that $ g(A) = Y(A) $ for some even
superderivation Y of \sca \ (to be called the
\emph{infinitesimal generator} of $
\Phi_t $).

From Eq.(3), we have, for small t,
\begin{eqnarray}
(\Phi_t)_* X \simeq X + t [Y, X] = X + t L_Y X.
\end{eqnarray}

\noindent \textbf{Proposition 2.7} \emph{Given $\Phi_t$ and Y as
above and a p-form $ \omega, $ we have, for small t,}
\begin{eqnarray}
\Phi_t^* \omega \simeq \omega - t L_Y \omega.
\end{eqnarray}
\emph{Proof.} We have
\begin{eqnarray*}
(\phstup_t \omega)(X_1,..,X_p)
& = & \Phi_t^{-1} [\omega\left( (\Phi_t)_* X_1,..(\Phi_t)_* X_p \right)] \\
& \simeq & \omega(X_1,..,X_p) - tY\omega(X_1,..,X_p)  \\
    & \ & + t \sum _{i=1}^p \omega(X_1,.., [Y,X_i],.., X_p) \\
& = & [\omega - t L_Y \omega] ](X_1,..,X_p).  \hspace{.2in} \Box
\end{eqnarray*}

\vspace{.12in} \noindent \textbf{2.4 A generalization of the DVNCG
scheme}

\vspace{.1in} It is easily seen that, in the developments in the
last two subsections, it is possible to restrict the
superderivations to a (topological) Lie sub-superalgebra
\scx \ of \sdera \ which is also a Z(\sca)-module and
develop the whole formalism with the pair $(\sca, \scx)$ obtaining
thereby a useful generalization of the superderivation-based
differential calculus. Working with such a pair is the analogue of
restricting oneself to a leaf of a foliated manifold as the example
below indicates.

\vspace{.1in} \noindent \emph{Example.}  $ \sca =
C^{\infty}(\mathbb{R}^3; \mathbb{R})$; \scx = the Lie subalgebra
of the real Lie
algebra $\scx(\mathbb{R}^3)$ of  vector fields on $\mathbb{R}^3$
generated by the Lie differential operators $ L_j =
\epsilon_{jkl}x_k\partial_l$ for the SO(3)-action on $\mathbb{R}^3$
(here $\partial_l \equiv \frac{\partial}{\partial x^l}$). In this case
Z(\sca) consists of constant functions [implying $Z(\sca) = \mathbb{R}$]
and \scx \ is obviously a Z(\sca)-module. The
differential operators $L_j$, when expressed in terms of the polar
coordinates $r, \theta, \phi$, contain only the partial derivatives
with respect to $\theta$ and $\phi$; they, therefore, act on the
2-dimensional spheres that constitute the leaves of the foliation
$\mathbb{R}^3 - \{(0,0,0) \} \simeq S^2 \times \mathbb{R}$. The
restriction [of the pair $(\sca, \scx(\mathbb{R}^3))$] to
(\sca,\scx) amounts to restricting oneself to a leaf ($S^2$) in the
present case.

\vspace{.1in} In the generalized formalism, one obtains the cochains
$C^{p,s}(\scx, \sca)$ for which the formulas of  sections  2.2 and
2.3 are valid (with the $X_js$ restricted to \scx). The
differential forms $\Omega^{p,s}(\sca)$ will now be replaced by the
objects $ \Omega^{p,s} (\scx, \sca)$ obtained by restricting the
cochains to the $Z(\sca)$-linear ones. [In the new notation, the
objects $\Omega^{p,s}(\sca)$ is $\Omega^{p,s}(\sdera,\sca)$.]

\vspace{.1in} To define the induced mappings \phst \ and \phstup \
in the present context, one must employ a (topological)
\emph{pair-isomorphism} $\Phi : (\sca,\scx) \rightarrow
(\scb, \scy)$ which consists of  a (topological)
superalgebra *- isomorphism $ \Phi : \sca \rightarrow \scb$ such
that the induced mapping $\Phi_* : \sdera \rightarrow SDer(\scb)$
restricts to a (topological) isomorphism of \scx \  onto \scy. The
properties of the induced mappings described in propositions 2.4
and 2.6 continue to hold.

Given a one-parameter family of transformations [i.e. (topological)
pair automorphisms]
$\Phi_t : (\sca, \scx) \rightarrow (\sca, \scx)$, the
condition $(\Phi_t)_* \scx \subset \scx $ implies that the
infinitesimal generator Y of $\Phi_t$ must satisfy the condition
$[Y,X] \in \scx $ for all $X \in \scx$. In practical applications
one will generally have $Y \in \scx$ which trivially satisfies this
condition.

This generalization will be used in sections 3.7 and 3.8 and in
later papers in the series.

\vspace{.12in} \noindent \textbf{2.5 Superderivations and
differential forms on tensor products of superalgebras}

\vspace{.12in} Given two superalgebras $\sca^{(1)}$ and
$\sca^{(2)}$, their algebraic (skew) tensor product $\sca =
\sca^{(1)} \otimes \sca^{(2)}$ has, as elements, finite sums of
tensored pairs :
\begin{eqnarray*}
\sum_{j=1}^{m} A_j \otimes B_j \ \ \ \  A_j \in \sca^{(1)}, \ \ B_j
\in \sca^{(2)};
\end{eqnarray*}
it is an algebra with the multiplication rule
(assuming $B_j$ and $A_k$ are homogeneous)
\begin{eqnarray*}
(\sum_{j=1}^{m} A_j \otimes B_j)(\sum_{k=1}^{n} A_k \otimes B_k) =
\sum_{j,k} \eta_{B_jA_k} (A_jA_k) \otimes (B_jB_k).
\end{eqnarray*}
This tensor product will be taken to be equipped with the
projective topology [53] [it is then denoted as
$ \sca^{(1)} \otimes_{\pi}
\sca^{(2)};$ its completion will be denoted as $ \sca^{(1)}
\hat{\otimes}_{\pi} \sca^{(2)}$ or simply $ \sca^{(1)}
\hat{\otimes} \sca^{(2)}.$]

\noindent \emph{Note.} Henceforth, we shall generally skip
explicit mention of the words `topological' and `continuous';
their use in appropriate situations should be understood.

The superalgebra $\sca^{(1)}$ (resp. $\sca^{(2)}$) has, in \sca, an
isomorphic copy consisting of the elements ($A \otimes I_2, A \in
\sca^{(1)}$) (resp. $I_1 \otimes B, B \in \sca^{(2)}$) to be denoted
as $\tilde{\sca}^{(1)}$ (resp. $\tilde{\sca}^{(2)}$) where $I_1$ and
$I_2$ are the unit elements of $\sca^{(1)}$ and $\sca^{(2)}$. We
shall also use the notations $\tilde{A}^{(1)} = A \otimes I_2$ and $
\tilde{B}^{(2)} = I_1 \otimes B$.

Superderivations and differential forms on $\sca^{(i)}$ and
$\tilde{\sca}^{(i)}$ (i = 1,2) are formally related through the
induced mappings corresponding to the isomorphisms $ \Xi^{(i)} :
\sca^{(i)} \rightarrow \tilde{\sca}^{(i)}$ given by $\Xi^{(1)}(A) =
A \otimes I_2$ and $\Xi^{(2)}(B) = I_1 \otimes B$. For example,
corresponding to $X \in SDer(\sca^{(1)}),$ we have $ \tilde{X}^{(1)}
= \Xi^{(1)}_*(X)$ in SDer($\tilde{\sca}^{(1)}$) given by [see
Eq.(3)]
\begin{eqnarray}
\tilde{X}^{(1)}(\tilde{A}^{(1)}) = \Xi^{(1)}_*(X)(\tilde{A}^{(1)}) =
\Xi^{(1)}[X(A)] = X(A) \otimes I_2.
\end{eqnarray}
Similarly, corresponding to $Y \in SDer(\sca^{(2)})$, we have
$\tilde{Y}^{(2)} \in \mbox{SDer}(\tilde{\sca}^{(2)})$ given by
$\tilde{Y}^{(2)}(\tilde{B}^{2}) = I_1 \otimes Y(B)$. For the 1-forms
$\alpha \in \Omega^1(\sca^{(1)})$ and $\beta \in
\Omega^1(\sca^{(2)})$, we have $\tilde{\alpha}^{(1)} \in
\Omega^1(\tilde{\sca}^{(1)})$ and $\tilde{\beta}^{(2)} \in
\Omega^1(\tilde{\sca}^{(2)})$ given by [see Eq.(12)]
\begin{eqnarray}
\tilde{\alpha}^{(1)}(\tilde{X}^{(1)}) = \Xi^{(1)}[\alpha
(((\Xi^{(1)})^{-1})_* \tilde{X}^{(1)})] = \Xi^{(1)}[\alpha(X)] =
\alpha(X) \otimes I_2
\end{eqnarray}
and $ \tilde{\beta}^{(2)}(\tilde{Y}^{(2)}) = I_1 \otimes \beta(Y)$.
Analogous formulas hold for the higher forms.

We can extend the action of the superderivations $ \tilde{X}^{(1)}
\in SDer(\tilde{\sca}^{(1)})$ and $\tilde{Y}^{(2)} \in
SDer(\tilde{\sca}^{(2)})$ to $\tilde{\sca}^{(2)}$ and
$\tilde{\sca}^{(1)}$ respectively by defining
\begin{eqnarray*}
\tilde{X}^{(1)}(\tilde{B}^{(2)}) = 0 , \mbox{\ \ } \tilde{Y}^{(2)}
(\tilde{A}^{(1)}) = 0 \mbox{\ \ for all \ } A \in \sca^{(1)} \mbox{\
and \ } B \in \sca^{(2)}.
\end{eqnarray*}
It is useful to note that
\[ A \otimes B = (A \otimes I_2)(I_1 \otimes B) = \tilde{A}^{(1)}
\tilde{B}^{(2)}; \hspace{1.2in} (19+) \]
hence, for homogeneous $X  \in \sdera$ \ and $A \in \sca^{(1)},$
\begin{eqnarray} X(A \otimes B) = X(\tilde{A}^{(1)} \tilde{B}^{(2)}) =
(X\tilde{A}^{(1)})\tilde{B}^{(2)} + \eta_{XA} \tilde{A}^{(1)}
X(\tilde{B}^{(2)}). \end{eqnarray} It follows that an $X \in \sdera$
is determined completely by its action on the subalgebras
$\tilde{\sca}^{(1)}$ and $\tilde{\sca}^{(2)}.$

\vspace{.1in} \noindent \textbf{Proposition 2.8} \emph{Given the
superalgebras $\sca^{(i)}, \tilde{\sca}^{(i)}$ (i = 1,2) and \sca \
as above, every $X \in \sdera$ can be uniquely expressed as $X_1 +
X_2$ where $X_i
\in \sdera $ (i= 1,2) satisfy the conditions \\
(i) $ X_1 = X $ on $ \tilde{\sca}^{(1)}$ and = 0 on
$\tilde{\sca}^{(2)}$; \\
(ii) $X_2 = X $ on $\tilde{\sca}^{(2)}$ and = 0 on
$\tilde{\sca}^{(1)}$.}

\vspace{.1in} \noindent \emph{Proof.} Clearly, it is adequate to
consider homogeneous superderivations. Since X, as a mapping on
\sca, is linear, it will give, when acting on a general element of
\sca, a sum of terms of the form of the right hand side of Eq.(20).
Now, defining $X_1$ and $X_2$ as indicated in the statement of the
theorem and taking $\epsilon_{X_1} = \epsilon_{X_2} =\epsilon_{X}$,
the action of $X_1 + X_2$ on $A \otimes B$ gives the right hand side
of (20). To see uniqueness, let $ X = X_1 + X_2 = X_1^{\prime} +
X_2^{\prime}.$ Putting $Y_1 = X_1 - X_1^{\prime}, Y_2  =
X_2 - X_2^{\prime},$ it is easily seen that both $Y_1$ and $Y_2$ give
zero when acting on $\tilde{\sca}^{(1)}$ and $\tilde{\sca}^{(2)}$,
hence on \sca. \ $\Box$

\vspace{.1in} With the extensions described above, we have available
to us superderivations belonging to the span of terms of the form
[see Eq.(18)]
\begin{eqnarray}
X = X^{(1)} \otimes I_2 + I_1 \otimes X^{(2)}.
\end{eqnarray}
Here $I_1$ and $I_2$ are to be understood as the linear mappings
$\mu_1 (I_1) = id_{\sca^{(1)}}$ and $\mu_2 (I_2) = id_{\sca^{(2)}}$
where $\mu_i$ is the multiplication operator for the superalgebra
$\sca^{(i)}$ (i= 1,2). Replacing $I_2$ and $I_1$ in Eq.(21) by
elements of $Z(\sca^{(2)})$ and $Z(\sca^{(1)})$ respectively, we
again obtain superderivations of \sca. We, therefore, have  the
space of superderivations
\begin{eqnarray}
[SDer(\sca^{(1)}) \otimes Z(\sca^{(2)})] \oplus [Z(\sca^{(1)})
\otimes SDer(\sca^{(2)})].
\end{eqnarray}
This space, however, is generally only a Lie sub-superalgebra of
SDer(\sca). For example, for $\sca^{(1)} = M_m(\mathbb{C})$ and
$\sca^{(2)} = M_n(\mathbb{C}) (m, n > 1)$, recalling that all the
derivations of these matrix algebras are inner and that their
centers consist of scalar multiples of the respective unit matrices,
we have the (complex) dimensions of Der$(\sca^{(1)})$, and
Der$(\sca^{(2)})$ respectively, $(m^2 -1)$ and $(n^2-1)$ [so that
the dimension of the space (22) is $m^2 + n^2 -2$ ] whereas that of
\dera \ is $(m^2n^2-1)$.

We shall need to employ (in section 4) a class of superderivations
more general than (22). To introduce that class, it is instructive
to obtain explicit representation(s) for a general derivation  of
the matrix algebra $M_m(\mathbb{C}) \otimes M_n(\mathbb{C})$. We
have
\begin{eqnarray}
[A \otimes B, C \otimes D]_- & = & AC \otimes BD - CA \otimes DB \\
                             & = & [A,C]_- \otimes \frac{1}{2} [B,D]_+ +
                       \frac{1}{2}[A,C]_+ \otimes [B,D]_-.
\end{eqnarray}
This gives, in obvious notation,
\begin{eqnarray}
D_{A \otimes B} \equiv [A \otimes B,.]_-
& = & A.(.) \otimes B_.(.) - (.).A \otimes (.).B \\
& = & D_A \otimes J_B^{(2)} + J_A^{(1)} \otimes D_B
\end{eqnarray}
where $J_B^{(2)}$ is the linear mapping on $\sca^{(2)}$ given by
\mbox{$J_B^{(2)} (D) = \frac{1}{2}[B,D]_+$} and a similar expression
for $J_A^{(1)}$ as a linear mapping on $\sca^{(1)}$.  Eq.(25) shows
that an expression for a derivation of the algebra
$ \sca = \sca^{(1)} \otimes
\sca^{(2)} $ need not explicitly contain those of $\sca^{(1)}$ and
$\sca^{(2)}$. The expression (26) is more useful for us; it is a
special case of the more general form
\begin{eqnarray}
X = X_1 \otimes \Psi_2 + \Psi_1 \otimes X_2
\end{eqnarray}
where $X_i \in SDer(\sca^{(i)})$ (i=1,2) and $ \Psi_i : \sca^{(i)}
\rightarrow \sca^{(i)}$ (i =1,2) are linear mappings. Our
constructions in section 4.1  will lead us to structures of the
form (27). It is important to note, however, that an expression of
the form (27) (which represents a linear mapping of \sca \ into
itself) need not always be a (super-)derivation as can be easily
checked. One should impose the condition (1) on such an expression
to obtain a (super-)derivation.

A straightforward procedure to obtain general  differential forms
and the exterior derivative on \sca \ is to obtain the graded
differential space $(\Omega(\sca),d)$  as the tensor product
(Greub [47]) of the graded differential spaces
$(\Omega(\sca^{(1)}), d_1)$ and $ (\Omega(\sca^{(2)}), d_2)$. A
differential form in $\Omega^{k,t} (\sca)$ is of the form [notation
: $\alpha_{ir}^{(a)} \in \Omega^{i,r}(\sca^{(a)})$, a = 1,2]
\begin{eqnarray*}
\alpha_{kt} \ = \sum_{\begin{array}{c} i+j=k \\
r+s= t \ mod(2) \end{array}} \alpha^{(1)}_{ir} \otimes
\alpha^{(2)}_{js}. \hspace{1.2in}  (27+) \end{eqnarray*} The d
operation on $\Omega(\sca)$ is given by [here $\alpha \in
\Omega^p(\sca^{(1)})$ and $ \beta \in \Omega(\sca^{(2)})] $
\begin{eqnarray}
d(\alpha \otimes \beta) = (d_1 \alpha) \otimes \beta + (-1)^p \alpha
\otimes d_2 \beta.
\end{eqnarray}

We shall employ (in section 3.8 and in later papers) exterior
products of two even 1-forms on $\sca = \sca^{(1)} \otimes \sca^{(2)}$.
From the definition of exterior product in section 2.2, we have, for
$\alpha , \beta \in \Omega^{1,0} (\sca)$ and homogeneous $X, Y \in
\sdera$,\begin{eqnarray*}
(\alpha \wedge \beta) (X, Y) = \alpha (X) \beta (Y) -
\eta_{XY} \alpha(Y) \beta (X).  \hspace{.6in} (28+)
\end{eqnarray*}
Superderivations of \sca \
employed in concrete cases will be restricted to the form (27).

\vspace{.15in} \noindent \textbf{\large 3. Noncommutative Symplectic
Geometry and Hamiltonian Mechanics}

\vspace{.12in} We shall now  present a  treatment of noncommutative
symplectic geometry (extending the treatment of noncommutative
symplectic structures by Dubois-Violette and coworkers mentioned
above  so as to include proper treatment of canonical
transformations, Lie group actions etc in the noncommutative
setting) and Hamiltonian mechanics along lines parallel to the
developments in classical symplectic geometry and Hamiltonian
mechanics (Arnold [5]; Aldaya and Azcarraga [2]; Woodhouse [78];
Abraham and Marsden [1]; Souriau [72]).

\vspace{.15in} \noindent \textbf{3.1 Symplectic structures; Poisson
brackets}

\vspace{.1in} \noindent \emph{Note}. The sign conventions  adopted
below are parallel to those of Woodhouse. This results in a (super-)
Poisson bracket which, when applied to classical Hamiltonian
mechanics, differs from the `usual' one by a minus sign. [See
Eq.(56).] The main virtue of the adopted conventions is that Eq.(33)
below has no unpleasant minus sign on the right.

\vspace{.12in} A \emph{symplectic structure} on a (topological)
superalgebra \sca \ is  a (continuous) 2- form $\omega$
(the \emph{symplectic form}) which is even,
closed and \emph{non-degenerate} in the sense that, for every $A \in
\sca$, there exists a unique superderivation $Y_A$ in \sdera \ such
that
\begin{eqnarray}
i_{Y_A} \omega = - dA.
\end{eqnarray}
The pair $(\sca,\omega)$ will be called a \emph{symplectic
superalgebra}. A symplectic structure is said to be \emph{exact} if
the symplectic form is exact ( $ \omega = d \theta$  for some 1-form
$ \theta$ called the \emph{symplectic potential}).

A \emph{symplectic mapping} from a symplectic superalgebra
$(\sca,\alpha)$ to another one $(\scb,\beta)$ is a superalgebra
(topological) isomorphism $\Phi : \sca \rightarrow \scb$ such that
$\Phi^*\beta =
\alpha.$ [If the symplectic structures involved are exact, one
requires a symplectic mapping to preserve the symplectic potential
under the pull-back action; Eq.(14) then guarantees the preservation
of the symplectic form.] A symplectic mapping from a symplectic
superalgebra onto itself will be called a \emph{canonical/symplectic
transformation}. The symplectic form and its exterior powers are
invariant under canonical transformations.

If $\Phi_t$ is a  one-parameter family of canonical transformations
(satisfying the conditions (i) and (ii) mentioned in section 2.3)
generated by $X\in \sdera, $ the condition $ \Phi_t^* \omega =
\omega$ implies, with Eq.(17),
\begin{eqnarray}
L_X \omega = 0.
\end{eqnarray}
The argument just presented gives Eq.(30) with X an even
superderivation. More generally, a superderivation X (even or odd or
inhomogeneous) satisfying Eq.(30) will be called a \emph{locally
Hamiltonian} superderivation. Eq.(11) and the condition $ d \omega
=0 $  imply that Eq.(30) is equivalent to the condition
\begin{eqnarray} d (i_X \omega) = 0.
\end{eqnarray}
A \emph{(globally) Hamiltonian superderivation} X is one for which
the form $i_X \omega$ is exact. In Eq.(29), the object $Y_A$
is the Hamiltonian superderivation
corresponding to $A \in \sca$. Note from Eq.(29) that $\epsilon(Y_A)
= \epsilon(A)$. In analogy with the commutative case, we have

\vspace{.1in} \noindent \textbf{Proposition 3.1.} \emph{The
supercommutator of two locally Hamiltonian superderivations is a
globally Hamiltonian superderivation.}

\vspace{.1in} \noindent \emph{Proof.} Given two locally Hamiltonian
superderivations X and Y, we have [recalling Eq.(11) and the fact
that $\omega$ is even]
\begin{eqnarray*}
  i_{[X,Y]} \omega & = & (L_Y \circ i_X - i_X \circ
L_Y) \omega
                                         \nonumber \\
                 & = &  (i_Y \circ d +
                       d \circ i_Y)(i_X \omega)
                       \nonumber
                  =    d (i_Y i_X \omega)
\end{eqnarray*}
which is exact. $\Box$

\noindent It follows that the locally Hamiltonian superderivations
constitute a Lie superalgebra in which the globally Hamiltonian
superderivations constitute an ideal.

The \emph{Poisson bracket} (PB) of two elements A and B of \sca \ is
defined as
\begin{eqnarray}
\{ A, B \} = \omega (Y_A, Y_B) = Y_A (B).
\end{eqnarray}
With A, B homogeneous, we have the super-analogue of the
antisymmetry condition : $\{ A, B \} = - \eta_{AB} \{ B, A \}$ and
of the Leibnitz rule :
\begin{eqnarray*}
\{A, BC \} = Y_A (BC) & = & Y_A(B) C + \eta_{AB} B Y_A (C) \nonumber \\
                      & = & \{ A, B \} C + \eta_{AB} B \{ A, C \}.
\end{eqnarray*}
As in the classical case, we have the relation
\begin{eqnarray}
[Y_A, Y_B] = Y_{ \{A,B \} }.
\end{eqnarray}
Eqn.(33) follows by using the equation for $ i_{[X,Y]} \omega$ above
with $X = Y_A$ and $ Y =Y_B$ and equations (32) and (29),
remembering that Eq.(29) determines $Y_A$ uniquely. The super-Jacobi
identity
\begin{eqnarray}
0 & =  & \frac{1}{2} (d \omega)(Y_A, Y_B, Y_C)  \nonumber \\
  & =  & \{ A, \{ B, C \} \} + (-1)^{\epsilon_A(\epsilon_B + \epsilon_C)}
            \{ B,  \{ C, A \} \}  \nonumber \\
  & \, &    + (-1)^{\epsilon_C (\epsilon_A  + \epsilon_B)}
  \{ C, \{ A, B \} \}
\end{eqnarray}
is obtained by using Eq.(10) and noting that
\begin{eqnarray*}
Y_A[\omega(Y_B,Y_C)] & = & \{ A, \{B, C \} \} \\
\omega ([Y_A,Y_B],Y_C) & = &  \omega (Y_{\{A,B \}},Y_C) = \{ \{A,B
\}, C \}.
\end{eqnarray*}
Clearly, the pair (\sca, \{ ,  \}) is a Lie superalgebra. Eq.(33)
shows that the linear mapping $ A \mapsto Y_A $ is a
Lie-superalgebra homomorphism.

An element A of \sca \ can act, via $ Y_A $, as the infinitesimal
generator of a  one-parameter family of canonical transformations.
The change in $ B \in \sca$ \ due to such an infinitesimal
transformation is
\begin{eqnarray}
\delta B = \epsilon Y_A(B) = \epsilon \{ A, B \}.
\end{eqnarray}
In particular, if $ \delta B = \epsilon I$ (infinitesimal
`translation' in B), we have
\begin{eqnarray}
\{ A, B \} = I
\end{eqnarray}
which is the noncommutative analogue of the classical PB relation $
\{ p,q \} = 1$. A pair (A,B) of elements of \sca \ satisfying the
condition (36) will be called a \emph{canonical pair}.

\vspace{.1in} \noindent \emph{Note.} In general, one has a
generator of the form
$Y = Y_A$ when the action of the 1-parameter family $\Phi_t$ on
the system algebra \sca \ is `hamiltonian' as defined in section
3.5 below. In the treatment of dynamics in section 3.4 below,
such an action (on the system algebra) of the 1-parameter family
of transformations describing system evolution is implicitly
assumed. This is in keeping with the general practice (to be
followed, in close analogy with classical mechanics, in this
series of papers) to have the action of symmetry groups on
system algebras  hamiltonian. This serves to provide the
traditional correspondence between one-parameter families of
symmetry operations  and observables.

\vspace{.15in} \noindent \textbf{3.2 Reality properties of the
symplectic form and the Poisson bracket}

In the papers of Dubois-Violette and coworkers mentioned above,
the symplectic form has been taken to be real. It is desirable
to have a rationale for it, especially in the superalgebraic case.
In this subsection, we provide this rationale and obtain some
related results about Hamiltonian superderivations and Poisson
brackets which extend the corresponding results in
[35] to the superalgebraic case.

\vspace{.12in} For classical superdynamical systems, conventions
about reality properties of the symplectic form are based on the
fact that the Lagrangian is a real, even object (Berezin and Marinov
[8]; Dass [23]). The matrix of the symplectic form is then
real-antisymmetric in the `bosonic sector' and imaginary-symmetric
in the `fermionic sector' (which means anti-Hermitian in both
sectors). Keeping this in view, it appears appropriate to impose, in
noncommutative Hamiltonian mechanics, the following reality
condition on the symplectic form $ \omega$:
\begin{eqnarray}
\omega^* (X,Y) = - \eta_{XY}\omega (Y,X) \hspace{.2in} \mbox{for all
homogeneous} \ X, Y \in \sdera;
\end{eqnarray}
but this means, by Eq.(7), that $ \omega^* = \omega$ (i.e. $\omega$
is real) which is the most natural assumption to make about
$\omega$. Eq.(37) is equivalent to the condition
\begin{eqnarray*}
\omega(X^*,Y^*) = - \eta_{XY} [\omega (Y,X)]^*.
\end{eqnarray*}

\noindent \textbf{Proposition 3.2.} \emph{In a symplectic
superalgebra $(\sca, \omega)$ with $\omega$ real, we have, for
arbitrary $ A,B \in \sca,$ }
\begin{eqnarray} (i) \ (Y_A)^* = Y_{A^*}; \ \ (ii) \
\{ A,B \}^* = \{ A^*, B^* \}.
\end{eqnarray}

\noindent \emph{Proof.} (i) Eq.(29) gives, for any $X \in \sdera,$
\begin{eqnarray*}
- (dA)(X) = \omega(Y_A,X) = [\omega (Y_A^*, X^*)]^*,\end{eqnarray*}
which, on applying the *-operation, gives
\begin{eqnarray*}
(i_{Y_A^*} \omega)(X^*) = -[dA(X)]^* = - dA^*(X^*) \end{eqnarray*}
which, in turn, gives $i_{Y_A^*} \omega = -d A^*$ implying the
desired result.

\noindent (ii) We have
\begin{eqnarray*}
\{ A,B \}^* & = & [\omega(Y_A, Y_B)]^* = \omega^*(Y_A^*,Y_B^*)  \\
            & = & \omega (Y_{A^*}, Y_{B^*}) = \{A^*, B^* \}. \ \Box
\end{eqnarray*}

Eq.(38)(ii) is consistent with the  reality properties of the
classical and quantum Poisson brackets.[See equations (56) and (43)
below.] Note that the PB of two hermitian elements is hermitian.

\vspace{.12in} \noindent \textbf{3.3 Special superalgebras; the canonical
symplectic form}

\vspace{.1in} In this subsection, we shall consider a distinguished
class of superalgebras  which have a canonical symplectic structure
associated with them. As we shall see below and in paper II, these
superalgebras play an important role in quantum mechanics.

A complex, associative, non-supercommutative  superalgebra will be
called \emph{special} if all its superderivations are inner. The
differential 2-form $\omega_c$ defined on such a superalgebra \sca \
by
\begin{eqnarray}
\omega_c (D_A, D_B) = [A,B] \ \textnormal{for all} \ A,B \in \sca
\end{eqnarray}
is said to be the \emph{canonical form} on \sca. (\emph{Note.} This
definition of the canonical form differs from that of
Dubois-Violette [35] by a factor of i.) It is easily seen to be
closed [the equation $(d \omega_c)(D_A,D_B,D_C) = 0$ is nothing but
the Jacobi identity for the supercommutator], imaginary (i.e.
$\omega_c^* = - \omega_c$) and dimensionless. For any $A \in \sca$,
the equation \begin{eqnarray*} \omega_c(Y_A, D_B) = - (dA)(D_B) =
[A,B] \ \textnormal{for all} \ B \in \sca
\end{eqnarray*}
has the unique solution $ Y_A = D_A$. (To see this, note that, since
all superderivations are inner, $Y_A = D_{C}$ for some $C \in \sca$;
the condition $\omega_c (D_C, D_B) \  = [C,B] = [A,B]$ for all $B
\in \sca$ \ implies that $(C - A) \in Z(\sca)$. But then $D_C = D_A.
\ \Box$) This shows that $\omega_c$ is nondegenerate and we have
\begin{eqnarray}
i_{D_A} \omega_c = - dA.
\end{eqnarray}
The closed and non-degenerate form $\omega_c$ defines, on \sca, the
\emph{canonical symplectic structure}. It gives, as Poisson bracket,
the supercommutator :
\begin{eqnarray}
\{ A,B \} = Y_A(B) = D_A(B) = [A,B].
\end{eqnarray}
Using equations (40) and (11), it is easily seen that the form
$\omega_c$ is \emph{invariant} in the sense that $ L_X \omega_c = 0$
for all $X \in \sdera$. The invariant symplectic structure on the
algebra $M_n(C)$ of complex $n \times n$ matrices obtained by
Dubois-Violette and coworkers [37] is a special case of the invariant
canonical symplectic structure on special superalgebras described
above.

If, for a special superalgebra \sca, we take, instead of $\omega_c$,
 $\omega = b \ \omega_c$ as the symplectic form (where $b$ is
a nonzero complex number), we have
\begin{eqnarray}
Y_A = b^{-1} D_A ,  \hspace{.4in} \{A, B \} = b^{-1}[A,B].
\end{eqnarray}
The \emph{ quantum Poisson bracket}
 \begin{eqnarray} \{ A, B \}_Q = (-i \hbar)^{-1} [A,B]
 \end{eqnarray}
is a special case of this with $b = -i \hbar.$ (Note that b must be
imaginary to make $\omega$ real.) In the case of the
Schr$\ddot{o}$dinger representation for a nonrelativistic spinless
particle, the Heisenberg-Schr$\ddot{o}$dinger algebra $\sca_Q$
generated by the position and momentum operators $X_j, P_j$ (j=
1,2,3) {defined on the invariant dense domain
$\scs (\mathbb{R}^3)$ in the Hilbert space
$\sch = L^2(\mathbb{R}^3,dx)$ is special [35,36]; one has,
therefore, a canonical form $\omega_c$ and the \emph{quantum
symplectic structure}  on $\sca_Q$  given by the \emph{quantum
symplectic form}
\begin{eqnarray} \omega_Q = -i \hbar \omega_c \end{eqnarray}
which gives the PB (43).

We shall refer to  a symplectic structure of the above sort with a
general nonzero b  as the \emph{quantum symplectic structure with
parameter b}.

\vspace{.12in} \noindent \textbf{3.4 Noncommutative Hamiltonian
mechanics}

\vspace{.12in} We shall now present the formalism of NHM combining
elements of noncommutative symplectic geometry and noncommutative
probability in the algebraic framework developed above.

 \vspace{.1in} \noindent \textbf{\small 3.4.1 The system algebra
 and states}

\vspace{.1in} In NHM, one associates, with every physical system, a
symplectic superalgebra $(\sca, \omega)$. Here we shall treat the
term `physical system' informally as is traditionally done; some
formalities in this connection will be taken care of in paper II
where a provisional set of  axioms will be  given. The even
Hermitian elements of \sca \ will be identified as
\emph{observables} of the system. The collection of all observables
in \sca \ will be denoted as $\mathcal{O}(\sca)$; it is a real
linear space  closed under PBs and, therefore, constitutes a real
Lie subalgebra of the complex Lie super-algebra ($\sca, \{, \}$).

To take care of limit processes and continuity of mappings, we must
employ topological algebras. The choice of the admissible class of
topological algebras must meet the following reasonable
requirements:

\vspace{.1in} \noindent (i) It should be closed under the formation
of (a) topological completions and (b) tensor products. (Both are
nontrivial requirements [33].)

\noindent (ii) It should include

\noindent (a) the Op$^{*}$-algebras (Horuzhy [55]) based on the
pairs $(\sch, \scd)$ where \sch \ is a complex separable Hilbert
space and \scd \ a dense linear subset of \sch; [Such an algebra is
an algebra of operators which, along with their adjoints, map \scd \
into itself. The
*-operation on the algebra is defined as the restriction of the
Hilbert space adjoint on $\mathcal{D}$. These are the algebras of
operators (not necessarily bounded) appearing in the traditional
Hilbert space QM; for example, the operator algebra $\sca_Q$ in the
previous subsection  belongs to this class.]

\noindent (b) algebras of smooth functions on manifolds (to
accommodate classical Hamiltonian mechanics and permit a transparent
treatment of quantum-classical correspondence).

\noindent (iii) The GNS representations of the system algebra
induced by various states must have \emph{separable} Hilbert spaces
as the representation spaces.

\vspace{.1in} The right choice appears to be, as mentioned in
section 2.1, the $\hat{\otimes}$-(star-) algebras of Helemskii [53]
(i.e. locally convex *-algebras which are complete and Hausdorff
with a jointly continuous product) satisfying the additional
condition of being separable.  Henceforth (super-)algebras employed
in the treatment of NHM will generally be assumed to belong to this
class. Any additional requirements will be explicitly stated
when needed in later work.

\vspace{.1in} \noindent \emph{Note.} We do not adopt the
topological setting of
Iguri and Castagnino [56] (nuclear, barrelled $b^*$-algebras) for
the following reasons :

\noindent (i) The algebra of observables in the non-relativistic
quantum mechanics of a finite number of particles is not barrelled.
(Dubin and Hennings [33], p.97).

\noindent (ii) These authors go for barreledness to have a ``nice''
spectral theory. In our case, however, for noncommutative system
algebras, we shall see the `inevitability' of the traditional Hilbert
space theory (paper II). There is, therefore, nothing special to be
gained by going into the spectral theory at the algebraic level.

\noindent (iii) In paper II, we shall have a concrete realization
of the Dirac bra-ket formalism in its rigorous version in terms of
nuclear rigging of the quantum mechanical Hilbert space without
postulating nuclearity at the algebraic level.

\noindent (iv) The general attitude adopted by the author in
this context is to keep as much generality as possible at
the foundational level and introduce extra postulates, if
necessary, when dealing with quantum field dynamics.

\vspace{.1in}  A \emph{state} on a system algebra \sca \  is a
continuous linear
functional $\phi$ on \sca \ which is (i) positive [which means $
\phi(A^*A) \geq 0$ for all $A \in \sca$) and (ii) normalized [i.e.
$\phi (I) =1$]. Given a state $\phi$, the quantity $\phi(A)$ for any
observable A is real (this can be seen by considering, for example,
the quantity$\phi[(I + A)^*(I + A)]$) and is to be interpreted as
the expectation value of A in the state $\phi$. Following general
usage in literature, we shall call observables of the form $A^*A$ or
a sum of such terms \emph{positive} (strictly speaking, the term
`non-negative' would be more appropriate); states assign
non-negative expectation values to such observables. The family of
all states on \sca \ will be denoted as $\mathcal{S}(\sca)$; it
will be understood to inherit the subspace topology of the strong
dual $\sca_s^{\prime}$ (or  $\sca_b^{\prime}$) (Yosida [79],
section IV.7). It is closed under convex combinations: given $ \phi_i \in
\mathcal{S}(\sca)$ , i = 1,..,n and $p_i > 0 $ with $p_1 + ..+ p_n =
1$, we have $ \phi = \sum_{i=1}^n p_i \phi_i $ also in
$\mathcal{S}(\sca)$. States which cannot be expressed as nontrivial
convex combinations of other states will be called \emph{pure}
states and others \emph{mixed} states. The family of pure states of
\sca \  will be denoted as \sone(\sca). The triple $(\sca,
\sone(\sca), \omega)$ will be referred to as a \emph{symplectic
triple}. This is the proper noncommutative analogue of a symplectic
manifold $(M, \omega_{cl})$ in classical Hamiltonian mechanics. [
Note that specification of the phase space M serves to define both
the algebra $C^{\infty}(M)$ as well as its pure states which are
points of M.]

Expectation values of all even elements of \sca \ can be expressed
in terms of those of the observables (by considering the breakup of
such an element into its Hermitian and anti-Hermitian part). This
leaves out the odd elements of \sca. It appears reasonable to demand
that the expectation values $\phi(A)$ of all odd elements $ A \in
\sca$ must vanish for all  pure states (and, therefore, for all
states).

Denoting the (topological) dual of the superalgebra \sca \ by
$\sca^*$, a (topological)
automorphism $\Phi : \sca \rightarrow \sca$ induces the transpose
mapping $\tilde{\Phi} : \sca^* \rightarrow \sca^*$ such that
\begin{eqnarray}
\tilde{\Phi}(\phi)(A) = \phi(\Phi(A)) \ \mbox{or} \ <
\tilde{\Phi}(\phi), A> \ = \ <\phi, \Phi(A)>
\end{eqnarray}
where $<,>$ denotes the dual space pairing.  The mapping
$\tilde{\Phi}$ [which is easily seen to be linear
and bijective (and continuous)] maps states (which form
a subset of $\sca^*$) onto
states. To see this, note that

\vspace{.1in} \noindent (i) $ \tilde{\Phi}(\phi)(A^*A) =
\phi(\Phi(A^*A)) = \phi(\Phi(A^*) \Phi(A))
= \phi(\Phi(A)^*\Phi(A)) \geq 0;$

\vspace{.1in} \noindent (ii) $ [\tilde{\Phi}(\phi)](I) =
\phi(\Phi(I)) =\phi(I) = 1. $

\vspace{.1in} \noindent The linearity of $\tilde{\Phi}$ (as a
mapping on  $\sca^*$  ) ensures that it preserves convex
combinations of states. In particular, it maps pure states onto pure
states. We have, therefore, a bijective mapping $\tilde{\Phi} :
\sone(\sca) \rightarrow \sone(\sca).$ An automorphism of the set
\scs(\sca) \ of states is an invertible mapping $\Psi : \scs(\sca)
\rightarrow \scs(\sca)$ preserving convex combinations. The mapping
$\tilde{\Phi}$ is, therefore, an automorphism of the set
$\scs(\sca)$ of states restricting to a bijection on the set
$\sone(\sca)$ of pure states.

When $\Phi$ is a canonical transformation, the condition  $\phstup
\omega = \omega$ gives, for $X,Y \in \sdera$,
\begin{eqnarray*}
\omega(X,Y) = (\phstup \omega)(X,Y) = \Phi^{-1}[\omega(\Phi_*X,
\phst Y)]
\end{eqnarray*}
which gives
\begin{eqnarray}
\Phi[\omega(X,Y)] = \omega (\phst X, \phst Y).
\end{eqnarray}
Taking expectation value of both sides of this equation in a state
$\phi$, we get
\begin{eqnarray}
(\tilde{\Phi}\phi)[\omega(X,Y)] = \phi[\omega(\phst X, \phst Y)].
\end{eqnarray}
 When $\Phi$ is an infinitesimal canonical transformation generated
by $G \in \sca$, we have
\begin{eqnarray*}
\tilde{\Phi}(\phi) (A) = \phi (\Phi(A)) \simeq \phi (A + \epsilon \{
G,A \}).
\end{eqnarray*}
Putting $\tilde{\Phi}(\phi) = \phi + \delta \phi$, we have
\begin{eqnarray}
(\delta \phi)(A) = \epsilon \phi (\{ G,A \}).
\end{eqnarray}

\vspace{.1in} \noindent \textbf{\small 3.4.2 Dynamics}

\vspace{.1in} Dynamics of the system is described in terms of the
one-parameter family $\Phi_t$ of canonical transformations generated
by an observable H, called the \emph{Hamiltonian}. (The parameter t
is supposed to be an evolution parameter which need not always be
the conventional time.) Writing $\Phi_t(A) = A(t)$ and recalling
Eq.(35), we have the \emph{Hamilton's equation} of NHM :
\begin{eqnarray}
\frac{dA(t)}{dt} = Y_H[A(t)] = \{ H, A(t) \}.
\end{eqnarray}
The triple $(\sca, \omega, H)$ [or, more appropriately, the
quadruple $(\sca, \sone (\sca), \omega, H)$] will be called an
\emph{NHM Hamiltonian system} or simply a \emph{noncommutative
Hamiltonian system}. It is the analogue of a classical Hamiltonian
system $ (M, \omega_{cl}, H_{cl})$ [where (M, \omcl) is a symplectic
manifold and $H_{cl}$, the classical Hamiltonian (a smooth
real-valued function on M)]. As far as the evolution is concerned,
the Hamiltonian is, as in the classical case, arbitrary up to the
addition of a constant multiple of the unit element. We shall assume
that H is bounded below in the sense that its expectation values in
all pure states (hence in all states) are bounded below.

This is the analogue of the Heisenberg picture in traditional QM. An
equivalent description, the analogue of the Schr$\ddot{o}$dinger
picture, is obtained by transferring the time dependence to states
through the relation [see Eq.(45)]
 \begin{eqnarray}< \phi(t), A> = <\phi, A(t)> \end{eqnarray}
where $\phi(t) = \tilde{\Phi}_t(\phi)$. The mapping $\tilde{\Phi}_t$
on states satisfies the condition (47) [with $\Phi = \Phi_t$]
which may be
said to represent the canonicality of the evolution of states. With
$ \Phi = \Phi_t$ and G = H, Eq.(48) gives the \emph{Liouville
equation} of NHM:
\begin{eqnarray}
\frac{d \phi (t)}{dt}(A) = \phi (t)( \{ H,A \}) \hspace{.12in}
\mbox{or} \hspace{.12in} \frac{d \phi(t)}{dt} (.) = \phi (t)(\{ H,.
\}).
\end{eqnarray}

\vspace{.1in} \noindent \textbf{\small 3.4.3 Equivalent
descriptions; Symmetries and conservation laws}

\vspace{.1in}
 By a `description' of a system, we shall mean specification of its
 triple $(\sca, \scs(\sca), \omega)$. Two descriptions are said to
 be \emph{equivalent} if they are
related through a pair of automorphisms $\Phi_1 : \sca \rightarrow
\sca$  and $\Phi_2 : \scs(\sca) \rightarrow \scs(\sca)$ such that
the symplectic form and the expectation values are preserved :
\begin{eqnarray}
\Phi_1^* \omega = \omega; \hspace{.12in} \Phi_2 (\phi)[\Phi_1(A)] =
\phi(A)
\end{eqnarray}
for all $ A \in \sca $ and $\phi \in \scs(\sca)$. The second
equation above and Eq(45) imply that we must have $\Phi_2 =
(\tilde{\Phi}_1)^{-1}$. Two equivalent descriptions are, therefore,
related through a canonical transformation on \sca \ and the
corresponding inverse transpose transformation on the states. An
infinitesimal transformation of this type generated by $ G \in \sca$
takes the form [see equations (35) and (48)]
\begin{eqnarray}
\delta A = \epsilon \{ G, A \}, \hspace{.2in} (\delta \phi)(A) = -
\epsilon \phi (\{ G,A \})
\end{eqnarray}
for all $A \in \sca$ and $ \phi \in \scs (\sca).$

These transformations may be called symmetries of the formalism;
they are the analogues of simultaneous unitary transformations on
operators and state vectors in a Hilbert space preserving
expectation values of operators. Symmetries of dynamics are the
subclass of these which leave the Hamiltonian invariant:
\begin{eqnarray}
\Phi_1 (H) = H.
\end{eqnarray}
For an infinitesimal transformation generated by $ G \in \sca $,
this equation gives
\begin{eqnarray}
\{ G,H \} = 0.
\end{eqnarray}
It now follows from the Hamilton's equation (49) that (in the
`Heisenberg picture' evolution) G is a constant of motion. This is
the situation familiar from classical and quantum mechanics:
generators of symmetries of the Hamiltonian are conserved quantities
and vice-versa.

\vspace{.1in} \noindent \emph{Note}. Redundancy in the definition of
symmetry operations given above permits some flexibility in their
implementation. It is often useful to implement them such that they
act, in a single implementation, \emph{either} on observables
\emph{or} on states, and the two actions are related as the
Heisenberg and Schr$\ddot{o}$dinger picture evolutions above [see
equations (50) and (45)]; we shall refer to this type of
implementation as \emph{unimodal}. In such an implementation, the
second equation in (53) will not have a minus sign on the right.

\vspace{.1in} For future reference, we define equivalence of NHM
Hamiltonian systems. Two NHM Hamiltonian systems  $(\sca, \sone
(\sca), \omega, H)$ and \linebreak $(\sca^{\prime}, \sone
(\sca^{\prime}), \omega^{\prime}, H^{\prime})$  are said to be
equivalent if they are related through a pair $\Phi = (\Phi_1,
\Phi_2)$ of homeomorphisms such that $\Phi_1 : \sca
\rightarrow \sca^{\prime}$ is a symplectic
mapping connecting the Hamiltonians [i.e. $ \Phi_1^* \omega^{\prime}
= \omega$ and $\Phi_1 (H) = H^{\prime}]$ and $\Phi_2: \sone(\sca)
\rightarrow \sone(\sca^{\prime})$ such that $<\Phi_2(\phi),
\Phi_1(A)> \  =  \ <\phi, A>$ for all $A \in \sca$ and $\phi \in
\sone(\sca).$

\vspace{.15in} \noindent \textbf{\small 3.4.4 Classical Hamiltonian
mechanics and traditional Hilbert space QM as subdisciplines of NHM}

\vspace{.1in} A classical hamiltonian system $(M, \omega_{cl},
H_{cl})$ is a special case of an NHM Hamiltonian system $(\sca,
\sone(\sca), \omega, H)$ with $\sca = \sca_{cl} \equiv
C^{\infty}(M, \mathbb{R}), \ \sone(\sca) = M, \ \omega = \omega_{cl} \equiv \sum
dp_j \wedge dq^j$ (in canonical coordinates) and H = $H_{cl}$; the
NHM PBs are, in this case, the classical PBs
\begin{eqnarray} \{ f,g \}_{cl} = \sum_{j=1}^{n}\left(\frac{\partial f}
{\partial p_j}\frac{\partial g}{\partial q_j}- \frac{\partial
f}{\partial q_j} \frac{\partial g}{\partial p_j}\right).
\end{eqnarray}
Eq.(49) now becomes the classical Hamilton's equation. Representing
states by probability densities in the phase space M, Eq.(51) goes over,
in appropriate cases (for $ M = R^{2n}$, for example, after the
obvious partial integrations)  to the classical Liouville equation
for the density function.

To see the traditional Hilbert space QM as a subdiscipline of NHM,
it is useful to introduce the concept of a \emph{quantum triple}
$(\sch, \mathcal{D}, \sca)$ where \sch \ is a complex separable
Hilbert space, $\mathcal{D}$ a dense linear subset of \sch \ and
\sca \ an Op$^*$-algebra of operators based on ($\sch,\mathcal{D}$).
Here we shall consider only the \emph{standard quantum triples} by
which we mean those in which (i) the algebra \sca \ is special in
the sense of section 3.3, and (ii) \sca \ acts irreducibly on
($\sch,\mathcal{D}$) [i.e. there does not exist a smaller quantum
triple $(\sch^{\prime}, \mathcal{D}^{\prime}, \sca)$ with
$\mathcal{D}^{\prime} \subset \mathcal{D}, \ \sca
\mathcal{D}^{\prime} \subset \mathcal{D}^{\prime}$ and
$\sch^{\prime}$ is a proper subspace of \sch]. More general
situations involving superselection rules will be covered in the
systematic treatment of quantum systems in paper II.

With \sca \ special, one can define the quantum symplectic form
$\omega_Q$ as in Eq.(44); the pair $(\sca, \omega_Q)$ may be called
a \emph{quantum symplectic algebra}. With \sca-action irreducible,
the space $\sone(\sca)$ of pure states of \sca \ consists of vector
states corresponding to normalized vectors in $\mathcal{D}$.
Choosing an appropriate self adjoint element H of \sca \ as the
Hamiltonian operator, we have a quantum hamiltonian system $(\sca,
\sone(\sca), \omega_Q, H)$  as a special case of an NHM Hamiltonian
system. With the quantum PBs of Eq.(43), Eq.(49) goes over to the
traditional Heisenberg equation of motion. General states are
represented by density operators $\rho$ satisfying the condition
$|Tr(\overline{\rho A})| < \infty$ for all  $A \in \sca$ where the
overbar indicates closure ([33]; p. 121). Noting that
$Tr(\overline{\rho_1 A}) = Tr (\overline{\rho_2 A})$ for all $A \in
\sca$ implies $\rho_1 = \rho_2$, Eq.(51) goes over to the von
Neumann equation
\begin{eqnarray} \frac{d\rho(t)}{dt} = (-i\hbar)^{-1}[\rho(t), H].
\end{eqnarray}

\vspace{.12in} \noindent \textbf{3.5 Symplectic actions of Lie
groups in NHM}

\vspace{.1in} The study of symplectic actions of  Lie groups in NHM
proceeds generally parallel to the classical case (Sudarshan and
Mukunda [73]; Alonso [3]; Arnold [5]; Guillemin and Sternberg [49];
Woodhouse [78]) and promises to be quite rich and rewarding. Here we
shall present the essential developments mainly to provide
background material for sections 3.6 and 3.8 and paper II.

Let G be a connected Lie group with Lie algebra \scg. Elements of G,
\scg \ and \gstar \  (the dual space of \scg) will be denoted,
respectively, as g,h,.., $\xi, \eta,..$ and $\lambda, \mu,..$. The
pairing between \gstar \ and \scg \ will be denoted as $<.,.>$.
Choosing a basis $ \{\xi_a; a = 1,..,r\}$ in \scg, we have the
commutation relations $ [ \xi_a, \xi_b] = C_{ab}^c \xi_c.$ The dual
basis in \gstar \ is denoted as $ \{ \lambda^a \}$ (so that $ <
\lambda^a, \xi_b> = \delta^a_b$). The action of G on \scg \ (adjoint
representation) will be denoted as $ Ad_g : \scg \rightarrow \scg$
and that on \gstar \ (the coadjoint representation) by $Cad_g :
\gstar \rightarrow \gstar;$ the two are related as $ < Cad_g
\lambda, \xi> = <\lambda, Ad_{g^{-1}}\xi>.$ With the bases chosen as
above, the matrices in the two representations are related as $
(Cad_g)_{ab} = (Ad_{g^{-1}})_{ba}.$

Recalling the mappings $\Phi_1$ and $\Phi_2$ of the previous
subsection, a \emph{symplectic action} of of G on a symplectic
superalgebra $(\sca, \omega)$ is given by the assignment, to each $
g \in G,$ a symplectic mapping
$\Phi_1(g) : \sca \rightarrow \sca$ which is a group action [which
means that $\Phi_1(g) \Phi_1(h) = \Phi_1(gh)$ and $\Phi_1(e)=
id_{\sca}$ in obvious notation]. The group action is assumed to
have continuity properties so as to permit construction of
infinitesimal generators of 1-parameter subgroups (the objects
$Z_{\xi}$ below) as in section 2.3. The action on the states is
given by the mappings $ \Phi_2(g) = [\tilde{\Phi}_1(g)]^{-1}$.

A one-parameter subgroup g(t) of G generated by $\xi \in \scg$
induces a locally Hamiltonian superderivation $Z_{\xi} \in
SDer(\sca)$ as the generator of the one-parameter family
$\Phi_1(g(t)^{-1})$ of canonical transformations of \sca.  For
small t
\begin{eqnarray*} \Phi_1(g(t)^{-1})(A) \simeq A + t Z_{\xi}
(A). \end{eqnarray*} [\emph{Note.} We employed $\Phi_1(g(t)^{-1})$
(and not $ \Phi_1(g(t))$) for defining $Z_{\xi}$ above because the
former correspond to a right action of G on \sca.]

\vspace{.1in} \noindent \textbf{Proposition 3.3.} \emph{The
correspondence $ \xi \rightarrow Z_{\xi}$ is a Lie algebra
homomorphism} :
\begin{eqnarray}  Z_{[\xi,\eta]} = [Z_{\xi}, Z_{\eta}].
\end{eqnarray} A proof of (58), whose steps are parallel to those
for Lie group actions on manifolds (Matsushima [61]; Dass [23]), was
given in the unpublished paper [27] [it is an instructive
application of the mathematical
techniques of section 2; in particular, the induced mappings \phst
\ of Eq.(3) play a role analogous to that of the mappings on vector
fields induced by diffeomorphisms]; we shall skip the details here.

A symplectic G-action is said to be \emph{hamiltonian} if the
superderivations $Z_{\xi}$ are Hamiltonian, i.e. for each $\xi \in
\mathcal{G},\ Z_{\xi} = Y_{h_{\xi}}$  for some $h_{\xi} \in \sca$
(called the \emph{hamiltonian} corresponding to $\xi$). These
hamiltonians are arbitrary up to  addition of  multiples of the unit
element. This arbitrariness can be somewhat reduced by insisting
that $h_{\xi}$ be linear in $\xi$. (This can be achieved by first
defining the hamiltonians for the members of a basis in
$\mathcal{G}$ and then for general elements as appropriate linear
combinations of these.) We shall always assume this linearity.

A hamiltonian G-action satisfying the additional condition
\begin{eqnarray}
\{ h_{\xi}, h_{\eta} \} = h_{[\xi,\eta]} \ \mbox{for all \ } \xi,
\eta \in \mathcal{G}
\end{eqnarray}
is called a \emph{Poisson action}. Note that (recalling the
statement after proposition 3.2) the hamiltonians of a Poisson
action can be consistently chosen to be observables.

The following proposition is a straightforward noncommutative
generalization of the corresponding result in classical
Hamiltonian mechanics

\vspace{.1in} \noindent \textbf{Proposition 3.4.} \emph{The
hamiltonians of the Poisson action of a connected Lie group G on a
symplectic superalgebra $(\sca, \omega)$ have the following
equivariance property}:
\begin{eqnarray}
\Phi_1(g) (h_{\xi}) = h_{Ad_g(\xi)} \ \ \forall g \in G \ \
\textnormal{and} \ \ \xi \in \scg.
\end{eqnarray}
\emph{Proof.} Since G is connected, it is adequate to verify this
relation for infinitesimal group actions. Denoting by g(t) the
one-parameter group generated by $\eta \in \mathcal{G}$, we have,
for small t,
\begin{eqnarray*}
\Phi_1(g(t))(h_{\xi}) \hspace{.1in}   \simeq \hspace{.1in}  h_{\xi}
+ t \{ h_{\eta}, h_{\xi} \} \hspace{.1in}
                         = \hspace{.1in}  h_{\xi} + t h_{[\eta, \xi]}
         \hspace{.1in}  = \hspace{.1in}  h_{\xi + t [\eta, \xi]}
           \hspace{.1in} \simeq \hspace{.1in}  h_{Ad_{g(t)}\xi}
 \end{eqnarray*}
completing the verification. $\Box$

A Poisson action is not always admissible. The obstruction to such
an  action is determined by the objects
\begin{eqnarray}
\alpha (\xi, \eta) = \{h_{\xi}, h_{\eta} \} - h_{[\xi, \eta]} \ \
\xi, \eta \in \scg
\end{eqnarray}
which are easily seen to have vanishing Hamiltonian derivations :
\begin{eqnarray*} Y_{\alpha(\xi,\eta)} & = & [Y_{h_{\xi}},
Y_{h_{\eta }}] - Y_{h_{[\xi, \eta]}}  =  [Z_{\xi}, Z_{\eta}] -
Z_{[\xi,\eta]} = 0 \end{eqnarray*} and hence vanishing Poisson
brackets with all elements of \sca. [This last condition defines the
so-called  \emph{neutral elements} [73] of the Lie algebra $(\sca,
\{, \})$. They clearly form a  vector space which will be denoted as
$\mathcal{N}$.] We also have
\begin{eqnarray*}
\alpha([\xi, \eta], \zeta) + \alpha([\eta, \zeta], \xi) + \alpha
([\zeta,\xi], \eta) = 0 \ \textnormal{for all} \ \xi, \eta, \zeta
\in \mathcal{G}.
\end{eqnarray*}
The derivation (Woodhouse [78]; p.44) of this result in classical
mechanics employs only the standard properties of PBs and remains
valid in NHM. Comparing this equation with Eq.(6), we see that $
\alpha(.,.) \in Z^2_0(\scg,\mathcal{N})$. A redefinition of the
hamiltonians $ h_{\xi} \rightarrow h^{\prime}_{\xi} = h_{\xi} +
k_{\xi}I$ (where the scalars $k_{\xi}$ are linear in $\xi$) changes
$\alpha$ by a coboundary term [see the first equation in (6)]:
\begin{eqnarray*}
\alpha^{\prime} (\xi,\eta) = \alpha(\xi, \eta) - k_{[\xi,\eta]}I
\end{eqnarray*}
showing that the obstruction is characterized by a cohomology class
of $\mathcal{G}$ [i.e. an element of $H^2_0(\scg,\mathcal{N})$]. It
follows that a
necessary and sufficient condition for the admissibility of Poisson
action of G on \sca \ is that it should be possible to transform
away all the obstruction 2-cocycles by redefining the hamiltonians,
or, equivalently, $H_0^2(\mathcal{G},\mathcal{N}) = 0.$

We now restrict ourselves to the special case, relevant for
application in  paper II (in the treatment of elemetary systems), in
which the cocycles $\alpha$ are multiples of the unit element
:\begin{eqnarray} \alpha(\xi, \eta) = \underline{\alpha}(\xi, \eta)
I.
\end{eqnarray}
Assuming that the hamiltonians $h_{\xi}$ are observables, the
quantities $\underline{\alpha}(\xi, \eta)$ must be real numbers.
This implies $\mathcal{N} = \mathbb{R}$. In this case, the relevant
cohomology group $H^2_0(\scg,\mathbb{R})$ is a real finite
dimensional vector space; we shall take it to be $\mathbb{R}^m$. In
this case, as in classical Hamiltonian mechanics [73,14,3],
hamiltonian group actions (more generally, Lie algebra actions) with
nontrivial neutral elements can be treated as Poisson actions of a
(Lie group with a) larger Lie algebra $\hat{\scg}$ obtained as
follows: Let $ \eta_r(.,.)(r = 1,..,m)$ be a set of representatives
in $Z^2_0(\scg,\mathbb{R})$ of a basis in $H^2_0(\scg,\mathbb{R})$.
We add extra generators $M_r $ (r = 1,..,m) to the basis $\{ \xi_a
\}$ of \scg \ and take the commutation relations of the larger Lie
algebra $\hat{\scg}$ as
\begin{eqnarray}
[\xi_a, \xi_b] = C_{ab}^c \xi_c + \sum_{r=1}^{m}\eta_r(\xi_a, \xi_b)
M_r; \ \
  [\xi_a, M_r] \ = \ 0 \ = \ [M_r, M_s].
\end{eqnarray}
The connected and simply connected Lie group $\hat{G}$ with the Lie
algebra $\hat{\scg}$ is called the \emph{projective group} of G
(called `projective covering group' of G by Cari$\tilde{n}$ena and
Santander [14]; we follow the terminology of Alonso [3]); it is
generally a central extension of the universal covering group
$\tilde{G}$ of G.

The hamiltonian action of G with the cocycle $\alpha$ now becomes a
Poisson action of $\hat{G}$ with the Poisson bracket relations
(writing $h_{\xi_a} = h_a, \ h_{M_r}= h_r$)
\begin{eqnarray} \{ h_a, h_b \} = C_{ab}^c h_c + \sum _{r=1}^{m}
\eta_r(\xi_a, \xi_b) h_r; \   \{h_a, h_r\} = 0 = \{h_r, h_s\}.
\end{eqnarray}

\vspace{.12in} \noindent \textbf{3.6 The noncommutative momentum map}.

\vspace{.1in} In classical mechanics, given a Poisson action of a
connected Lie group G on a symplectic manifold  $(M, \omega_{cl})$
[with hamiltonians $ h^{(cl)}_{\xi} \in C^{\infty}(M)$], a useful
construction is the so-called \emph{momentum map} (Souriau [72];
Arnold [5]; Guillemin and Sternberg [49]) $ P: M \rightarrow
\mathcal{G}^*$ given by
\begin{eqnarray}
< P(x), \xi>\mbox{\  = \ }h^{(cl)}_{\xi}(x) \ \ \forall x \in M
\mbox{\ and \ } \xi \in \mathcal{G}.
\end{eqnarray}
This map relates the symplectic action $\Phi_g$ of G on M ($\Phi_g :
M \rightarrow M, \Phi_g^* \omega_{cl} = \omega_{cl}\  \forall g \in
G$) and the transposed adjoint action on $\mathcal{G}^*$ through the
equivariance property
\begin{eqnarray}
P(\Phi_g(x)) = \widetilde{Ad_g}(P(x)) \ \forall x \in M \mbox{\ and
\ } g \in G.
\end{eqnarray}

Noting that points of M are pure states of the algebra $ \sca_{cl} =
C^{\infty}(M, \mathbb{R})$, the map P may be considered as the restriction to M
of the dual/transpose $\tilde{h}^{(cl)}: \sca_{cl}^* \rightarrow
\mathcal{G}^*$ (where $\sca_{cl}^*$ is the dual space of
$\sca_{cl}$) of the linear map $h^{(cl)}: \mathcal{G} \rightarrow
\sca_{cl}$ given by $ h^{(cl)}(\xi) = h^{(cl)}_{\xi}$:
\begin{eqnarray*}
< \tilde{h}^{(cl)}(u), \xi> = <u, h^{(cl)}(\xi)> \ \forall \ u \in
\mathcal{A}_{cl}^* \mbox{\ and \ } \xi \in \mathcal{G}.
\end{eqnarray*}
The analogue of M in NHM is $\sone = \sone(\sca)$. Defining $ h :
\mathcal{G} \rightarrow \sca $ by $ h(\xi) = h_{\xi},$ the analogue
of the momentum map in NHM is the mapping $\tilde{h}: \sone
\rightarrow \mathcal{G}^* $ (considered as the restriction to \sone
\ of the mapping $\tilde{h} : \sca^* \rightarrow \mathcal{G}^*$)
given by
\begin{eqnarray}
<\tilde{h}(\phi), \xi> = < \phi, h(\xi) > = < \phi, h_{\xi}> \
\textnormal {for all}\  \phi \in \sone \ \textnormal{and}\  \xi \in
\scg.
\end{eqnarray}

\noindent \textbf{Proposition 3.5.} \emph{The noncommutative
momentum map $\tilde{h}$ has the following equivariance property: In
the notation employed above}
\begin{eqnarray} \tilde{h}(\Phi_2(g) \phi)
= Cad_g(\tilde{h}(\phi)) \  \textnormal {for all} \ \phi \in \sone \
\textnormal{and}\  g \in G.
\end{eqnarray}
\emph{Proof.} We have, for any $\xi \in \scg$,
\begin{eqnarray*}
< \tilde{h}(\Phi_2(g) \phi ), \xi> & = &  <\Phi_2(g) \phi, h_{\xi}>
                         =  < \phi, \Phi_1 (g^{-1})(h_{\xi})>
                         =  < \phi, h_{Ad_{g^{-1}}(\xi)} >  \\
                       & = & <\phi, h(Ad_{g^{-1}}(\xi))>
                   = < Cad_g (\tilde{h}(\phi)), \xi>
\end{eqnarray*}
giving Eq.(68). $\Box$

\vspace{.1in} \noindent \emph{Note}. In Eq.(68),the co-adjoint (and
not the transposed adjoint) action appears on the right because
$\Phi_2(g)$ is inverse transpose (and not transpose) of $\Phi_1(g)$.
With this understanding, (66) is obviously a special case of (68).

We shall make use of the noncommutative momentum map in the
treatment of elementary systems in paper II.

\vspace{.12in} \noindent \textbf{3.7 Generalized symplectic
structures and Hamiltonian systems}

\vspace{.12in} The generalization of the DVNCG scheme introduced in
section 2.4 can be employed to obtain the corresponding
generalization of the NHM formalism. One picks up a distinguished
Lie sub-superalgebra \scx \ of \sdera \ and restricts the
superderivations of \sca \ in all definitions and constructions to
those in \scx. Thus, a symplectic superalgebra $(\sca, \omega)$ is
now replaced by a \emph{generalized symplectic superalgebra} $(\sca,
\scx, \omega)$ and a symplectic mappings $\Phi : (\sca, \scx,
\alpha) \rightarrow (\scb, \scy, \beta)$ is restricted to a (topological)
superalgebra-isomorphism  $\Phi : \sca \rightarrow \scb$  such that
$\phst : \scx \rightarrow \scy$ is a Lie-superalgebra- isomorphism
and $\phstup \beta = \alpha$. An NHM  Hamiltonian system $(\sca,
\sone(\sca), \omega, H)$ is now replace by a \emph{generalized NHM
Hamiltonian sytem} $(\sca, \sone(\sca), \scx, \omega, H)$.

We shall use the generalized symplectic structures in the next
subsection and in paper II where we shall employ the pairs (\sca,
\scx) with \scx = ISDer(\sca) to define quantum symplectic structure
on superalgebras admitting outer as well as inner superderivations.

\vspace{.12in} \noindent \textbf{3.8 Augmented symplectics including
`time'; the noncommutative analogue of Poincar$\acute{e}$-Cartan
form}

\vspace{.1in} We shall now augment the symplectic kinematics
of NHM by including the evolution parameter t (`time') and develop the
non-commutative analogue of classical time dependent Hamiltonian
mechanics [1,2,78].

For a system S with associated symplectic superalgebra $ (\sca,
\omega)$ we construct the \emph{extended system algebra} $\sca^e =
C^{\infty}(\mathbb{R}) \otimes \sca $ (where the real line
$\mathbb{R}$ is the carrier space of  `time' t) whose elements are
finite sums $\sum_i f_i \otimes A_i $ (with $f_i \in
C^{\infty}(\mathbb{R}) \equiv \sca_0$) which may be written as
$\sum_i f_i A_i $. This algebra is the analogue of the algebra of
functions on the `evolution space' of Souriau  (the Cartesian
product of the time axis and the phase space
--- often referred to as the phase bundle). The superscript e in
$\sca^e$, may, therefore, also be taken to refer  to `evolution'.

  Derivations on $\sca_0$ are of the form $g(t)\frac{d}{dt}$ and
one-forms of the form h(t)dt where g and h are smooth functions;
there are no nonzero higher order forms. We have, of course, $ dt
(\frac{d}{dt}) = 1$.

  Employing the mapping $ \Xi : \sca \rightarrow \sca^e$ given by
$ \Xi (A) = 1 \otimes A$, which  is an isomorphism of the algebra
\sca \ onto the subalgebra $\tilde{\sca} \equiv 1 \otimes \sca $ of
$\sca^e$, we write, for a p-form $\alpha$ on \sca, $ (\Xi^{-1})^*
(\alpha) = \tilde{\alpha}$ and extend this form on $\tilde{\sca}$ to
one on $\sca^e$ by defining $\tilde{\alpha}(\frac{d}{dt}, ...) = 0.$
Similarly, the mapping $\Xi_0 : \sca_0 \rightarrow \sca^e$ given by
$\sca_0 (f) = f \otimes I_{\sca}$ is an isomorphism of $\sca_0$ onto
the subalgebra $\tilde{\sca}_0 = \sca_0 \otimes I_{\sca}.$  Keeping
the notation  dt for the 1-form $\tilde{dt} \equiv
(\Xi_0^{-1})^* (dt)$ on  $\tilde{\sca}_0$, we may extend it to one on
$\sca^e$ by defining (dt)(X) =0 for all $X \in SDer(\tilde{\sca})$.

  The symplectic structure $\omega$ on \sca \ induces, on $\sca^e$,  a
generalized symplectic structure (of the type introduced in the
previous subsection) with the distinguished Lie sub-superalgebra
\scx \ of $SDer(\sca^e)$ taken to be the one consisting of the
objects $\{id_{\sca_0} \otimes D; D\in \sdera \}$ which constitute a
Lie sub-superalgebra of $SDer(\sca^e)$ isomorphic to \sdera, thus
giving a generalized symplectic superalgebra $(\sca^e, \scx,
\tilde{\omega})$. The corresponding PBs on $\sca^e$ are trivial
extensions of those on \sca \ obtained by treating the `time' t as
an external parameter; this amounts to extending the
$\mathbb{C}$-linearity of PBs on \sca \ to what is essentially
$\sca_0$-linearity :
\begin{eqnarray*}
\{ fA + gB, hC \}_{\sca^e} = fh \{A,C \}_{\sca} + gh \{ B,C
\}_{\sca} \end{eqnarray*} where, for clarity, we have put subscripts
on the PBs to indicate the underlying superalgebras. We shall often
drop these subscripts; the underlying (super-)algebra will be clear
from the context.

To describe dynamics in $\sca^e$, we extend the one-parameter family
$\Phi_t$ of canonical transformations on \sca \ generated by a
Hamiltonian $H \in \sca$ to a one-parameter family $\Phi_t^e$ of
transformations on $\sca^e$ (which are `canonical' in a certain
sense, as we shall see below)  given by
\begin{eqnarray*} \Phi_t^e
(fA) \equiv (fA)(t) = f(t) A(t) \equiv (\Phi_t^{(0)} f) \Phi_t(A)
\end{eqnarray*} where $\Phi_t^{(0)}$ is the one-parameter group of
translations in `time' whose action on $\sca_0$ is given by
$(\Phi_t^{(0)} f) (s) = f(s + t)$; extension of the action of
$\Phi_t^e$ to general elements of $\sca^e$ is obtained by linearity.
An infinitesimal transformation under the evolution $\Phi_t^e$ is
given by
\begin{eqnarray*} \delta(fA)(t) & \equiv & (fA)(t + \delta t) -
(fA)(t) \\
& = & [\frac{df}{dt} A + f \{H, A \}_{\sca}] \delta t \equiv
\hat{Y}_H (fA) \delta t \end{eqnarray*} where \begin{eqnarray}
\hat{Y}_H = \frac{\partial} {\partial t} + \tilde{Y}_H.
\end{eqnarray} Here $\frac{\partial} {\partial t}$ is the derivation
on $\sca^e$ corresponding to the derivation $\frac{d}{dt}$ on
$\sca_0$ [i.e. $\frac{d}{dt} \otimes id_{\sca}$]
and [identifying  \sca \ with the  subalgebra $\tilde{\sca}$ of
$\sca^e$ to replace
$\{H, A \}_{\sca}$ by $\{ H, A \}_{\sca^e}$]
\begin{eqnarray} \tilde{Y}_H = \{ H, . \}_{\sca^e}.
\end{eqnarray} Note that

\noindent (i) $ dt(\hat{Y}_H ) = 1$;

\noindent (ii) the right hand side of Eq.(70) (where H should be
understood as $\tilde{H} = 1 \otimes H \in \sca^e$) remains well
defined if H is a general element of $\sca^e$ (`time dependent'
Hamiltonian). Henceforth, in this subsection, H will be understood
to be a general element of $\sca^e$.

Writing, for $F \in \sca^e, \Phi_t^e (F) = F(t),$ the obvious
generalization of the NHM Hamilton's equation (49) to $\sca^e$ is
the equation
\begin{eqnarray}
\frac{d F(t)}{dt} = \hat{Y}_H F(t) = \frac{\partial F(t)}{\partial
t} + \{ H(t), F(t) \}.
\end{eqnarray}

We next consider an object in $\sca^e$ which contains complete
information about the symplectic structure \emph{and} dynamics [i.e.
about $\tilde{\omega}$ and H (up to an additive constant multiple of
I)] and is determined by these objects. It is the 2-form
\begin{eqnarray}
\Omega =  \tilde{\omega} - dH  \wedge dt.
\end{eqnarray}
Here d is the exterior derivative in $\sca^e$ induced by the
exterior derivatives $d_1$ in $\sca_0$ and $d_2$ in \sca \ according
to Eq.(28) and  $dt = d_1 t \otimes I_{\sca}$ is the same as
$\tilde{dt}$ above. Since H and t are both even, the exterior product
on the right in Eq.(72) is of the form given by Eq.(28+). Noting that,
by Eq.(28), the form $\tilde{\omega} = 1 \otimes
\omega$ is closed, the form $\Omega$ is easily seen to be  closed.
If the symplectic structure on \sca \ is exact (with $ \omega = d_2
\theta$), we have [recalling Eq.(11++)] $ \Omega = d \Theta $ where
\begin{eqnarray}
\Theta = \tilde{\theta} - H dt
\end{eqnarray}
is the NHM analogue  of the Poincar$\acute{e}$-Cartan form in
classical mechanics.

Given $H = \sum_i f_i H_i$, we have
\[ dH = \sum_i (d_1 f_i) H_i + \sum_i f_i (d_2 H_i) \equiv
\tilde{d}_1 H + \tilde{d}_2 H. \] Note that \noindent (a) since
\begin{eqnarray*} \tilde{d}_1 H = \sum_i(d_1 f_i) H_i  =  \sum_i
(\frac{\partial f_i}{\partial t} dt) \otimes H_i  & = & \sum_i
(\frac{\partial f_i}{\partial t} \otimes H_i) (dt \otimes 1) \\ & =
& \frac{\partial H}{\partial t} dt,
\end{eqnarray*} we have $ dH \wedge dt = \tilde{d}_2 H \wedge dt$;

\noindent (b) since the symplectic superalgebra $(\tilde{\sca},
\tilde{\omega})$ is isomorphic to $(\sca, \omega)$ and since the
functions $f_i$ in H act essentially as numbers in the context of
the symplectic structure on $\tilde{\sca}$, we must have
\[ i_{\tilde{Y}_H} \tilde{\omega} = - \tilde{d}_2 H. \]

The closed form $\Omega$ is generally not non-degenerate. It defines
what may be called a \emph{presymplectic structure} (Souriau [72])
on $\sca^e$. In fact, we have here the noncommutative analogue of a
special type of presymplectic structure called \emph{contact
structure} [1,2,9]; it essentally
means that the presymplectic form is minimally degenerate.

A \emph{symplectic action} of a  Lie group G on the presymplectic
space $(\sca^e, \Omega)$ is the assignment, to every $g \in G$, an
automorphism $\Phi(g)$ of the superalgebra $\sca^e$ having the usual
group action properties and such that \\(i) $\Phi(g)^* \Omega =
\Omega$; \\ (ii) $\Phi(g) \tilde{\sca}_0 \subset \tilde{\sca}_0$
where $ \tilde{\sca}_0 = \sca_0 \otimes I$. \\ The condition (i)
implies, as in section 3.5, that, to every element $\xi$ of the
Lie algebra \scg \ of G, corresponds a derivation $\hat{Z}_{\xi}$ of
$\sca^e$ such that $L_{\hat{Z}_{\xi}} \Omega = 0$ which, in view of
the condition $d \Omega = 0$, is equivalent to the condition
\begin{eqnarray}
d(i_{\hat{Z}_{\xi}} \Omega) = 0.
\end{eqnarray}

We shall now verify that the one-parameter family $\Phi_t^{e}$ of
transformations on $\sca^e$ is symplectic/canonical. Time
translations employed in the description of dynamics clearly satisfy
the condition (ii). To verify (i), it is adequate to verify that
Eq.(74) holds with $\hat{Z}_{\xi} = \hat{Y}_H$. We have, in fact,
much more :

\vspace{.1in} \noindent \textbf{Proposition 3.6.} \emph{In the
notation employed above, we have}
\begin{eqnarray}
  i_{\hat{Y}_H} \Omega = 0.
\end{eqnarray}
\emph{Proof.} We have
\begin{eqnarray*}
i_{\hat{Y}_H} \Omega & = & i_{\partial/\partial t} \Omega +
                           i_{\tilde{Y}_H} \Omega \\
& = & - i_{\partial/\partial t}( \tilde{d}_2 H \wedge dt) +
                     i_{\tilde{Y}_H} \tilde{\omega} - i_{\tilde{Y}_H}
                     (\tilde{d}_2 H \wedge dt) \\
                       & = & \tilde{d}_2H - \tilde{d}_2H
                       -i_{\tilde{Y}_H}(\tilde{d}_2H) dt \\
               & = & [i_{\tilde{Y}_H}(i_{\tilde{Y}_H} \tilde{\omega})]dt =
               0. \ \ \Box
\end{eqnarray*}
Eq.(75) will be exploited below to prove the noncommutative
analogue of the classical symplectic version of Noether's theorem.

\vspace{.15in} \noindent \textbf{3.9 Noncommutative symplectic
version of Noether's theorem}

\vspace{.12in} The classical Noether's theorem [2,72,23] is
developed in the Lagrangian formalism which applies to systems
with configuration space. Its symplectic version is obtained
in the framework of time-dependent Hamiltonian mechanics in
the phase bundle $\tilde{P} = \mathbb{R} \times T^*M$  where M
is the configuration manifold; here the analogue of the Lagrangian
is the classical Poincar$\acute{e}$-Cartan form
\[ \Theta_{cl} = \tilde{\theta}_0 - H_{cl} dt \]
where $ \tilde{\theta}_0 = \pi_2^* \theta_0$ is the pull-back
on $\tilde{P}$ of of the canonicl 1-form $\theta_0 = p_j dq^j$
on the cotangent bundle $T^*M$ of M. The pair $(\tilde{P},
\Omega_{cl})$ where $\Omega_{cl} = d \Theta_{cl}$, constitutes a
contact manifold  (which means that $\Omega_{cl}$ is a closed
form of maximal rank on the odd dimensional manifold
$\tilde{P}$); it provides a more general representation of
a dynamical system than a traditional Hamiltonian system.

Let X be a vector field on $\tilde{P}$ generating a one
parameter family of local diffeomorphisms of $\tilde{P}$. This family
will be called a \emph{Noether symmetry} of the contact manifold
$(\tilde{P},\Omega_{cl})$ if, for some function $f_X$ on
$\tilde{P}$, we have
\[ \ \ \ \ L_X \Theta = - d f_X; \ \ \ \ \ \ \ \ \ (75+) \]
this condition is the phase space analogue of the invariance of
the Lagrangian up to a total time derivative. The \emph{Noether
invariant} corresponding to this Noether symmetry is
\[ \ \ \ \ F_X = i_X \Theta + f_X \ \ \ \ \ \ \ \  (75++) \]
which can be shown [2] to be conserved when the equations of
motion are satisfied. With $ L_X = i_X \circ d + d \circ i_X$,
Eq.(75+) gives
\[\ \ \ \ i_X \Omega_{cl} = - d F_X \ \ \ \ \ \ \ (75+++) \]
which shows [on comparison with Eq.(29); see also Eq.(76) below]
that $F_X$ is the `hamiltonian' corresponding to the
`hamiltonian action' of the Noether symmetry on the contact
manifold $(\tilde{P},\Omega{cl})$.

Eq.(75+++) permits a generalization to general symplectic
manifolds obtained by taking  $\tilde{P} = \mathbb{R} \times P$
where $(P, \omega_{cl})$ is a general (finite dimensional)
symplectic manifold and
\[ \  \Omega_{cl} = \pi_2^* \omega_{cl} - d H \wedge d t.
\ \ \ \ (75++++) \]
The Noether invariant $F_X$ is again conserved on dynamical
trajectories. We shall now obtain the noncommutative analogue of
this generalized symplectic version of Noether's theorem.

Coming back to our presymplectic space $(\sca^e, \Omega)$,
a symplectic G-action  is said to be
\emph{hamiltonian} if the 1-forms $i_{\hat{Z}_{\xi}} \Omega$ are
exact, i.e. to each $\xi \in \scg$, corresponds a `hamiltonian'
$\hat{h}_{\xi} \in \sca^e$ (unique up to an additive constant
multiple of the unit element) such that
\begin{eqnarray}
i_{\hat{Z}_{\xi}} \Omega = -d \hat{h}_{\xi}.
\end{eqnarray}

\noindent \textbf{Theorem (1)} (Noncommutative symplectic version of
Noether's theorem). \emph{Given the presymplectic space $(\sca^e,
\Omega)$ associated with the symplectic algebra $(\sca, \omega)$ and
the Hamiltonian H (an even hermitian element of $\sca^e$), the `hamiltonians' (\emph{Noether
invariants}) $\hat{h}_{\xi}$ corresponding to the Hamiltonian action
of a connected Lie group G as in Eq.(76) are constants of motion.}

\vspace{.1in} \noindent \emph{Proof.} We have \begin{eqnarray}
\frac{d \hat{h}_{\xi}(t)}{dt} & = & \hat{Y}_H(\hat{h}_{\xi}(t))
                                      = (d \hat{h}_{\xi})(\hat{Y}_H)(t)
                             \nonumber \\
                  & = & - (i_{\hat{Z}_{\xi}} \Omega)(\hat{Y}_H)(t)
                        = 0
\end{eqnarray}
where, in the last step, Eq.(75) has been used. $\Box$

Some concrete examples of noncommutative Noether invariants will
be given in paper II.

\vspace{.15in} \noindent \textbf{\large 4. Interacting Systems in
Noncommutative Hamiltonian Mechanics}

\vspace{.12in}  We shall now consider the interaction of two systems
$S_1$ and $S_2$ described individually as the NHM Hamiltonian
systems $ (\sca^{(i)}, \omega^{(i)}, H^{(i)})$ (i=1,2) and treat the
coupled system $S_1 + S_2$ also as an NHM  Hamiltonian system. To
facilitate this, we must obtain the relevant mathematical objects
for the coupled system. The system algebra for the coupled system
will be taken to be the (topological completion of) (skew,
projective) tensor product $\sca = \sca^{(1)}
\otimes \sca^{(2)}$. [\emph{Note.} Since topological completion is
not relevant for the discussion below, we shall ignore it.] We next
consider the symplectic structure on \sca.

\vspace{.12in} \noindent \textbf{4.1 The symplectic form and Poisson
bracket on the (skew) tensor product of two symplectic
superalgebras}

\vspace{.1in} We shall freely use the notations and constructions in
section 2.5.

Given the symplectic forms $\omega^{(i)}$ on $\sca^{(i)}$ [with PBs
$\{, \}_i$ (i=1,2)] we explore the possibility of constructing a
(preferably unique) \emph{canonically induced} symplectic form
$\omega$ on \sca. The term `canonically induced' means that the
construction of $\omega$ should use nothing besides the algebraic
and symplectic structures on the superalgebras $\sca^{(1)}$ and
$\sca^{(2)}$. [\emph{Note.} No fundamental principle is violated if
$\omega$ is taken to depend on objects representing the interaction
between the two systems. In the formalism of NHM as developed above,
however, the burden of the description of dynamics  is carried by
the Hamiltonian in the kinematic framework provided by  the
symplectic structure; we shall, therefore, not consider this more
general possibility in this paper.] To ensure this, we impose on
$\omega$ (apart from the obvious condition that it must be a
symplectic form) the following conditions :

\noindent (a) It should not depend on anything other than the
objects $\omega^{(i)}$ and $I_{(i)}$ (i=1,2). (Note that the unit
elements are the only distinguished elements of the algebras being
considered.)

\noindent (b) The restrictions of $\omega$ to $\tilde{\sca}^{(1)}$
and $\tilde{\sca}^{(2)}$ be, respectively, $\omega^{(1)} \otimes
I_2$ and $I_1 \otimes \omega^{(2)}$.

\noindent Recalling  Eq.(27+), these
requirements lead to the unique choice
\begin{eqnarray}
\omega =  \omega^{(1)} \otimes I_2 + I_1 \otimes \omega^{(2)}.
\end{eqnarray}
Note that [using the relation $(v \otimes w)^* = v^* \otimes w^*$
for the involution on the tensor product of two vector spaces
with given involutions] the reality of $\omega^{(1)}$ and
$\omega^{(2)}$ implies the reality of $\omega.$
 To verify that it is a symplectic form, we must show
that it is (i) closed and (ii) nondegenerate. Eq.(28) gives
\begin{eqnarray*}
d \omega = (d_1 \omega^{(1)}) \otimes I_2 + \omega^{(1)} \otimes d_2
(I_2) + d_1 (I_1) \otimes \omega^{(2)} + I_1 \otimes d_2
\omega^{(2)} = 0
\end{eqnarray*}
showing that $\omega$ is closed. It follows, therefore, that, on
\sca, a canonically induced presymplectic structure given by the
presymplectic form (78) always exists. The precise statement about the
symplectic structure and Poisson bracket on the tensor product algebra
\sca \ appears in theorem (2) below.The rest of this subsection is
devoted to proving this theorem.

To show the nondegeneracy of
$\omega$, it is necessary and sufficient to show that, given $A
\otimes B \in \sca$ (with A and B homogeneous), there exists a
unique homogeneous superderivation $Y = Y_{A\otimes B}$ in \sdera \
such that
\begin{eqnarray}
i_{Y} \omega  =  - d (A\otimes B)
             & = & - (d_1 A ) \otimes B - A \otimes d_2 B  \nonumber \\
             & = & i_{Y_A^{(1)}} \omega^{(1)} \otimes B
               + A \otimes i_{Y_B^{(2)}} \omega^{(2)}
\end{eqnarray}
where $Y_A^{(1)}$ and $Y_B^{(2)}$ are the Hamiltonian
superderivations associated with $A \in \sca^{(1)}$ and $B \in
\sca^{(2)}$ respectively.

In the following developments, we shall denote the multiplication
operators in $\sca^{(1)}, \sca^{(2)}$ and \sca \ by $\mu_1, \mu_2$
and $\mu$ respectively.

\vspace{.1in} \noindent \textbf{Proposition 4.1} \emph{A linear
mapping  Y of $\sca = \sca^{(1)} \otimes \sca^{(2)}$  into itself is
a homogeneous superderivation satisfying Eq.(79) if and only if (i)
it is expressible as}
\begin{eqnarray}
Y = Y_A^{(1)} \otimes \Psi_{B}^{(2)} + \Psi_A^{(1)} \otimes
Y_B^{(2)}
\end{eqnarray}
\emph{where $\Psi_A^{(1)}$ and $\Psi_B^{(2)}$ are linear mappings (on
$\sca^{(1)}$ and $\sca^{(2)}$ respectively into themselves)
satisfying the conditions $\Psi_A^{(1)}(I_1) = A$ and $
\Psi_B^{(2)}(I_2) = B$, and (ii) it satisfies the relation (for
homogeneous $C \in \sca^{(1)}$ and $D \in \sca^{(2)}$ )}
\begin{eqnarray}
Y \circ \mu(C \otimes D) - \eta_{Y, C \otimes D} \ \mu(C \otimes D)
\circ Y = \mu(Y(C \otimes D)). \end{eqnarray} \emph{Proof.}
\emph{If} : Equations (78) and (80) give
\[ i_Y \omega  =  i_{Y_A^{(1)}} \omega^{(1)} \otimes \Psi_B^{(2)}
                     (I_2) \\ + \Psi_A^{(1)}(I_1) \otimes i_{Y_B^{(2)}} \omega^{(2)}
               =  \textnormal{right hand side of}\  (79). \]
Eq.(81) guarantees that Y is a superderivation of \sca.

\noindent \emph{Only if} : To be a superderivation, Y must satisfy,
by proposition (2.1), Eq.(81). Let $ X \in \sdera$ such that $ i_X
\omega = i_Y \omega$ where Y now stands for the right hand side of
(80). By proposition 2.1, X satisfies Eq.(81). We, therefore, need
only show that $Z \equiv X - Y = 0.$ Let $Z_1, Z_2 \in \sdera$ such
that $Z = Z_1 + Z_2$ is the unique decomposition of Z in accordance
with proposition 2.8; it means that \\
(i) $Z_1$ equals Z on $\tilde{\sca}^{(1)}$ and 0 on
$\tilde{\sca}^{(2)}$; \\
(ii) $Z_2$ equals 0 on $\tilde{\sca}^{(1)}$ and Z on
$\tilde{\sca}^{(2)}$. \\
We shall now prove, exploiting the non-degeneracy of
$\tilde{\omega}^{(1)} = \omega^{(1)} \otimes I_2$ and
$\tilde{\omega}^{(2)} = I_1 \otimes \omega^{(2)}$ (as symplectic
forms on $\tilde{\sca}^{(1)}$ and $\tilde{\sca}^{(2)}$
respectively), that $Z_1 = 0 = Z_2$. We have \[ 0 = i_Z \omega = i_Z
(\tilde{\omega}^{(1)} + \tilde{\omega}^{(2)})
\] which gives
\[ i_{Z_1} \tilde{\omega}^{(1)} = i_Z \tilde{\omega}^{(1)} = - i_Z
\tilde{\omega}^{(2)} = - i_{Z_2} \tilde{\omega}^{(2)}. \] Now
$i_{Z_1} \tilde{\omega}^{(1)}$ is a one-form on $\tilde{\sca}^{(1)}$
(hence of the form $ \alpha \otimes I_2$ where $\alpha$ is a
one-form on $\sca^{(1)}$) whereas $- i_{Z_2} \tilde{\omega}^{(2)}$
is a one-form on $\tilde{\sca}^{(2)}$ (hence of the form $I_1
\otimes \beta$ where $\beta$ is a one-form on $\sca^{(2)}$); the two
can be equal (as forms on \sca) only when $\alpha = 0 = \beta$.
Nondegeneracy of $\tilde{\omega}^{(1)}$ and $\tilde{\omega}^{(2)}$
now implies $Z_1 = 0 = Z_2. \ \Box $

\vspace{.1in} The following developments explore the consequences of
equations (80) and (81) leading ultimately to theorem (2).

\vspace{.1in} Noting that $ \mu (C \otimes D) = \mu_1(C) \otimes
\mu_2(D)$, Eq.(81) with Y of Eq.(80) gives
\begin{eqnarray} \eta_{BC} \{ [Y_A^{(1)} \circ \mu_1(C)] \otimes
[\Psi_B^{(2)} \circ \mu_2(D)] + [\Psi_A^{(1)} \circ \mu_1(C)]
\otimes
Y_B^{(2)} \circ \mu_2(D)] \}  \nonumber \\
- (-1)^{\epsilon} \{ [\mu_1(C) \circ Y_A^{(1)}] \otimes [\mu_2(D)
\circ \Psi_B^{(2)}] + [\mu_1(C) \circ \Psi_A^{(1)}] \otimes
[\mu_2(D) \circ Y_B^{(2)}]
\} \nonumber \\
= \eta_{BC} [\mu_1( \{ A,C \}_1) \otimes \mu_2(\Psi_B^{(2)}(D)) +
\mu_1(\Psi_A^{(1)}(C)) \otimes \mu_2(\{B,D \}_2)] \ \ \
\end{eqnarray} where $ \epsilon \equiv \epsilon_A \epsilon_C +
\epsilon_B \epsilon_D + \epsilon_B \epsilon_C$ and we have used the
relations $Y_A^{(1)}(C) = \{ A, C \}_1$ and $ Y_B^{(2)}(D) = \{ B,D
\}_2$.

The objects  $Y_A^{(1)}$ and $Y_B^{(2)}$, being superderivations,
satisfy relations of the form (1) :
\begin{eqnarray} Y_A^{(1)} \circ \mu_1(C) - \eta_{AC} \mu_1(C) \circ
Y_A^{(1)} = \mu_1(Y_A^{(1)}(C)) = \mu_1(\{ A, C \}_1) \nonumber  \\
Y_B^{(2)} \circ \mu_2(D) - \eta_{BD} \mu_2(D) \circ Y_B^{(2)} =
\mu_2( \{ B,D \}_2). \end{eqnarray} Putting $D = I_2$ in Eq.(82), we
have [noting that $\mu_2(D) = \mu_2(I_2) = id_{\sca^{(2)}},$  and $
\{B, I_2 \}_2 = Y_B^{(2)}(I_2) = 0$]
\begin{eqnarray}
  [Y_A^{(1)} \circ  \mu_1(C)] & \otimes  & \Psi_B^{(2)}  +
       [\Psi_A^{(1)} \circ \mu_1(C)] \otimes Y_B^{(2)} \nonumber \\
 & \ &     - \eta_{AC} \{ [\mu_1(C) \circ Y_A^{(1)}] \otimes \Psi_B^{(2)} +
      [\mu_1(C) \circ \Psi_A^{(1)}] \otimes Y_B^{(2)} \} \nonumber \\
 & \  & = \mu_1(\{ A, C \}_1) \otimes \mu_2(B) \end{eqnarray}
 which, along with equations (83), gives
 \begin{eqnarray} \mu_1( \{ A,C \}_1) \otimes
[\Psi_B^{(2)} - \mu_2(B)] & = & \nonumber \\
- [\Psi_A^{(1)} \circ \mu_1(C) & - & \eta_{AC} \mu_1(C) \circ
\Psi_A^{(1)}] \otimes Y_B^{(2)}. \end{eqnarray} Similarly, putting $
C = I_1$ in Eq.(82), we get
\begin{eqnarray} [\Psi_A^{(1)} - \mu_1(A)] \otimes \mu_2( \{ B,D
\}_2) & = & \nonumber \\
 - Y_A^{(1)} \otimes [\Psi_B^{(2)} \circ \mu_2(D) & - & \eta_{BD}
\mu_2(D) \circ \Psi_B^{(2)}]. \end{eqnarray}
Equations (85) and (86)  hold
for arbitrary homogeneous elements $A, C \in \sca^{(1)}$ and $B, D
\in \sca^{(2)}$. Note, however, that, in Eq.(85), both sides are
linear in B  and do not involve $\epsilon_B$ explicitly; hence this
equation holds for a general element B of $\sca^{(2)}$. Similarly, in
Eq.(86), A can be taken to be a general element of $\sca^{(1)}$. Now,
a careful consideration of these two equations  gives the relations
(see appendix)
\begin{eqnarray}
\Psi_A^{(1)} - \mu_1(A) = \lambda_1 Y_A^{(1)} \ \textnormal{for all}
\ A \in \sca^{(1)} \end{eqnarray}
\begin{eqnarray} \Psi_B^{(2)} \circ \mu_2(D) - \eta_{BD}\  \mu_2(D)
\circ \Psi_B^{(2)} = - \lambda_1 \ \mu_2( \{ B,D \}_2)
\end{eqnarray} \begin{eqnarray} \Psi_B^{(2)} - \mu_2(B) = \lambda_2
Y_B^{(2)} \ \textnormal{for all} \  B \in \sca^{(2)}
\end{eqnarray} \begin{eqnarray} \Psi_A^{(1)} \circ
\mu_1(C) - \eta_{AC}\  \mu_1(C) \circ \Psi_A^{(1)} = - \lambda_2 \
\mu_1( \{A,C \}_1) \end{eqnarray} where $\lambda_1$ and $\lambda_2$
are complex numbers (the possibility of either or both being zero
included).

Substituting for $\Psi_A^{(1)}$ and $\Psi_B^{(2)}$ from equations
(87) and (89)  in Eq.(80), we get
\begin{eqnarray}
Y  \equiv Y_{A \otimes B} & = & Y_A^{(1)} \otimes [\mu_2(B) +
\lambda_2
        Y_B^{(2)}] + [\mu_1(A) + \lambda_1 Y_A^{(1)}] \otimes Y_B^{(2)}
        \nonumber \\
  & = & Y_A^{(1)} \otimes \mu_2(B) + \mu_1(A) \otimes Y_B^{(2)} +
(\lambda_1 + \lambda_2) Y_A^{(1)} \otimes Y_B^{(2)}.
\end{eqnarray}
Note that only the combination $(\lambda_1 + \lambda_2) \equiv
\lambda$ appears in Eq.(91). To have a unique Y, we must obtain an
equation fixing $\lambda$ in terms of given quantities.

Substituting for $\Psi_A^{(1)}$  from  Eq.(87)  into Eq.(90) and
using the first of equations (83), we have
\begin{eqnarray}
\lambda \mu_1( \{ A,C \}_1)  =  - \mu_1([A,C]) \ \ \textnormal{for
all} \ \ A,C \in \sca^{(1)}
\end{eqnarray}
[argument for the phrase `for all' in Eq.(92) being the same as for
the equations (87) and (89)] which implies
\begin{eqnarray}
\lambda  \{ A,C \}_1   =  - [A,C] + K_1(A,C)
\end{eqnarray}
where $K_1(A,C)$ is an element of $\sca^{(1)}$ such that
\[ K_1(A,C) E = 0  \ \textnormal{for all} \ E \in \sca^{(1)}. \]
Since $\sca^{(1)}$ is unital, we must have $K_1 = 0$. Proceeding
similarly with the equations (89) and (88), we have, therefore,
\begin{eqnarray}
\lambda  \{ A,C \}_1   =  - [A,C] \  \textnormal{for all}
\ \ A,C \in \sca^{(1)}\\
\lambda  \{ B,D \}_2  =  - [B,D] \ \ \textnormal{for all} \ \ B,D
\in \sca^{(2)}.
\end{eqnarray} We have not one but two equations of the type we have
been looking for. This is a signal for the emergence of nontrivial
conditions (for the desired symplectic structure on the tensor
product superalgebra to exist).

Let us consider these equations  for the various possible situations
(corresponding to whether or not one or both the superalgebras are
super-commutative) :

\vspace{.12in} \noindent (i) Let $\sca^{(1)}$ be supercommutative.
Since the PB coming from a symplectic structure must be  nontrivial
(i.e. not identically zero), Eq.(94) implies that $\lambda = 0.$
Eq.(95) then implies that $\sca^{(2)}$ must also be
super-commutative. It follows that \\
(a) when both the superalgebras $\sca^{(1)}$ and $\sca^{(2)}$ are
super-commutative, the unique Y is given by Eq.(91) with $\lambda (
= \lambda_1 + \lambda_2) = 0$; \\
(b) when one of them is supercommutative and the other is not, the
$\omega$ of Eq.(78) does not define a symplectic structure on \sca.

\vspace{.12in} \noindent (ii) Let the superalgebra $ \sca^{(1)}$ be
non-supercommutative. Eq.(94) then implies that $ \lambda \neq 0,$
which, along with Eq.(95) implies that the superalgebra $\sca^{(2)}$
is also non-supercommutative [which is also expected from (b)
above]. We now have
\begin{eqnarray} \{ A,C \}_1 = - \lambda^{-1} [A,C], \ \
\{ B, D \}_2 = - \lambda^{-1} [B,D]. \end{eqnarray} Equations (96)
imply that, when both the superalgebras are non- \linebreak
supercommutative, a `canonically induced' symplectic structure on
their (skew) tensor product exists if and only if each superalgebra
has a quantum symplectic structure with the \emph{same} parameter
$(- \lambda)$, i.e.
\begin{eqnarray} \omega^{(1)} = - \lambda \omega^{(1)}_c, \ \
\omega^{(2)} = - \lambda \omega^{(2)}_c \end{eqnarray} where
$\omega^{(i)}_c$ (i=1,2) are the canonical symplectic forms on the
two superalgebras. Reality of these symplectic forms requires
$\lambda$ to be pure imaginary ($\lambda = i h_0$ with $h_0$ real).
The traditional quantum symplectic structure is obtained with $ h_0
= \hbar$ [see Eq.(44)].

\vspace{.1in} In all the permitted cases, the PB on the superalgebra
$ \sca = \sca^{(1)} \otimes \sca^{(2)}$ is given by
\begin{eqnarray} \{ A \otimes B, C \otimes D \} = Y_{A \otimes B}(C
\otimes D)
 & =  & \eta_{BC}
[\{ A, C \}_1 \otimes BD + AC \otimes \{ B, D \}_2  \nonumber \\
 & \ & + \lambda \{ A, C \}_1 \otimes \{ B,D  \}_2 ] \end{eqnarray}
 where the parameter $ \lambda$ vanishes in
 the super-commutative case; in the non-supercommutative case, it is
 the universal parameter appearing in the symplectic forms (97).

Noting that, in both the permitted cases,
\begin{eqnarray*} \lambda \{A,C \}_1 \otimes \{ B, D \}_2 & = &
-[A,C] \otimes \{B,D \}_2 = - \{ A,C \}_1 \otimes [B,D] \nonumber \\
& = & -\frac{1}{2} [A,C] \otimes \{B,D \}_2 - \frac{1}{2} \{A,C \}_1
\otimes [B,D],  \end{eqnarray*} the PB of Eq.(98) can be written in
the more symmetric form

\begin{eqnarray}
       \{ A&\otimes&B  ,  C \ \otimes \ D \}   \nonumber \\
 & = &  \eta_{BC}\left[ \{ A,C \}_1 \otimes \frac{BD +
\eta_{BD}DB}{2} + \frac{AC + \eta_{AC} CA}{2} \otimes  \{ B,D \}_2
\right]. \end{eqnarray}

To close the argument, we must check whether or not we have
exhausted the contents of Eq.(82) (recall that we have used only the
two special cases of this equation corresponding to $D = I_2$ and $
C = I_1$). Let $Y = Y_1 + Y_2$ be the (unique) breakup of Y in
accordance with the proposition 2.9. Writing an equation analogous
to (19+) for $C \otimes D$ and recalling Eq.(20), it is easily seen
that, by putting $ D = I_2$ in
Eq.(82), one is, in fact, considering $Y_1$; similarly, putting $C =
I_1$ amounts to considering $Y_2$. Exploring the implications of
Eq.(82) for $Y_1$ and $Y_2$ amounts to doing the same for Y. This
equation, therefore, cannot have any additional implications for Y.

We have proved the following theorem.

\vspace{.1in} \noindent \textbf{Theorem (2)}
(Commutative-noncommutative non-interaction and universality of
quantum symplectic structure).  \emph{(a) Given two symplectic
superalgebras $(\sca^{(i)}, \omega^{(i)})$ (i=1,2) with real
symplectic forms, the (skew)
tensor product $ \sca = \sca^{(1)} \otimes \sca^{(2)}$ always admits
the `canonically induced' presymplectic structure given by the
2-form $\omega$ of Eq.(78); \\
(b) it is a symplectic structure (with $\omega$ real) if and only
if either (i) both the
superalgebras are supercommutative, or (ii) both are
non-supercommutative and have a  quantum symplectic structure (in
the sense of section 3.3) with a universal parameter
$b = -i h_0$ where $h_0 \in \mathbb{R}$. \\
(c) In both the cases the PB on \sca \  is given by Eq.(99).}

\vspace{.1in} \noindent \emph{Note.} (i) The theorem says nothing
about the sign of the universal real parameter $h_0$  (which will be
later identified with the Planck constant $\hbar$). Given a
symplectic superalgebra $(\sca, \omega)$ with \sca \
non-supercommutative and $\omega$ real, we have generally the
freedom to replace $\omega$ by - $\omega$; using this freedom,
if necessary, the real parameter $h_0$ may be taken to be positive.

\vspace{.1in} \noindent (ii) The two forms $\omega^{(i)}$(i=1,2) of
Eq.(97) represent genuine symplectic structures only if the
superalgebras $\sca^{(i)}$ (i=1,2) are `special' (i.e. have only
inner superderivations; see section 3.3). Since the initial
objects $\sca^{(i)}$ (i=1,2) were assumed to be symplectic
superalgebras, an implicit conclusion of the theorem is that, in the
non-supercommutative case, these superalgebras must be special. A more
general version of the theorem is obtained by replacing, in part
(a),  the symplectic superalgebras $(\sca^{(i)}, \omega^{(i)})$ by
generalized symplectic superalgebras $(\sca^{(i)}, \scx^{(i)},
\omega^{(i)})$ (i=1,2); then, part (b) (ii) will take the modified
form : `both are non-supercommutative with $\scx^{(i)} =
ISDer(\sca^{(i)})$ and have a generalized quantum symplectic
structure ... $h_0 \in \mathbb{R}$.'

\vspace{.1in} \noindent (iii) The non-super version of Eq.(99) was
[wrongly, not realizing that Y of Eq.(80) is not always a
(super-)derivation] put forward by the author as the PB  for a
tensor product of algebras in the general case in (Dass [25]).
M.J.W. Hall pointed out to the author (private communication) that
it does not satisfy Jacobi identity in some cases, as shown, for
example, in (Caro and Salcedo [15]). Revised calculations by the
author then led to the results presented above.

\vspace{.1in} \noindent (iv) After the first version of the present
work appeared in the arXiv [0909.4606 v1 (math-ph)], L.L. Salcedo
drew the author's attention (private communication)  to the paper
(Salcedo [66]) in which similar results are obtained in an algebraic
treatment of systems employing a finite number of q-p pairs.

\vspace{.12in} \noindent \textbf{4.2 Dynamics of interacting systems}

\vspace{.1in} Given the individual systems $S_1$ and $S_2$ as the
NHM Hamiltonian systems $(\sca^{(i)}, \omega^{(i)},H^{(i)})$ (i =
1,2) where the two superalgebras are either both supercommutative or
both non-supercommutative, the coupled system ($S_1 + S_2$) has
associated with it a symplectic superalgebra $(\sca, \omega)$ where
$\sca = \sca^{(1)} \otimes \sca^{(2)}$ and $\omega$ is given by
Eq.(78). We form an NHM Hamiltonian system $(\sca, \omega, H)$ with
the Hamiltonian H given by
\begin{eqnarray}
H = H^{(1)} \otimes I_2 + I_1 \otimes H^{(2)} + H_{int}
\end{eqnarray}
where the interaction Hamiltonian is generally of the form
\begin{eqnarray*}
H_{int} = \sum_{i=1}^{n} F_i \otimes G_i.
\end{eqnarray*}

The evolution (in the Heisenberg type picture) of an obsevable
$A(t)\otimes B(t)$ of the coupled system is governed by the NHM
Hamilton's equation
\begin{eqnarray}
\frac{d}{dt} [ A(t) \otimes B(t)] & = & \{ H, A(t) \otimes B(t)\} \nonumber \\
   & = & \{H^{(1)},A(t)\}_1 \otimes B(t)
                     + A(t) \otimes \{H^{(2)},B(t) \}_2
                     \nonumber \\
                    & \ &  + \{H_{int}, A(t) \otimes B(t) \}
\end{eqnarray}
where we have used equations (49),(100) and (99). In the
Schr$\ddot{o}$dinger type picture, the time evolution of states of
the coupled system is given by the NHM Liouville equation (51) with
the Hamiltonian of Eq.(100). In favorable situations, the NHM
evolution equations for observables and states may be written for
finite time intervals by using appropriate exponentiations of the
evolution operators in equations (49) and (51).

\vspace{.1in} The main lesson from this section is that \emph{all}
systems in nature \emph{whose interaction with other systems can be
talked about} must belong to only one of the two `worlds' : the
`commutative world' in which all system superalgebras are
super-commutative and the `noncommutative world' in which all system
superalgebras are non-supercommutative with a \emph{universal}
(generalized) quantum symplectic structure. In view of the familiar
inadequacy of the commutative/classical physics, the `real' world
must clearly be the noncommutative world; its systems (satisfying an
extra condition of mutual compatibility between observables and pure
states to be introduced in paper II) will be called \emph{quantum
systems}. The classical systems with commutative system algebras and
traditional symplectic structures will appear only in the
appropriately defined classical limit (or, more generally, in the
classical approximation) of quantum systems.

\vspace{.2in} \noindent \textbf{\large 5. Concluding Remarks}

\vspace{.12in}  The development of NHM presented above constitutes
the first concrete step towards the desired unification of
probability theory and physics. The use of the observable-state
language in an algebraic setting is, of course, not new; what is new
is the unification of noncommutative \emph{symplectic} geometry with
noncommutative probability in such a setting. As we shall see in the
next and later papers in this series, this makes possible a much
better use of the true potential of the algebraic formalism.

Emergence of a natural place for the Planck constant in the
formalism (Theorem 2)  indicates quite strongly that we are
on the right track.

The absence of quantum-classical interaction (theorem 2) is, in
fact, a non-problem in NHM (or its augmented version, supmech,
described in paper II). Given a quantum system $S_1$ and a classical
system $S_2$, one can treat them as quantum systems, work in NHM
employing the formalism of section 4.2 and then take the classical
(or semiclassical) approximation for $S_2$ [remaining all the time
in NHM ( recall that NHM has both quantum and classical mechanics as
special subdisciplines)]. We shall see this idea at work in the
treatment of quantum measurements in paper III.

The reader should note the caution exercised in framing the first
sentence in the last para of the previous section (recall the phrase
in italics). Theorem (2) permits the construction of an NHM-based
formalism in which the universe as a whole is described `classically'
but all its subsystems whose interactions with other systems can be
talked about (i.e. meaningfully considered) are described quantum
mechanically.

The qualification in use of the term `classically' above is meant
to make the  point that an NHM Hamiltonian system with a
supercommutative system algebra need not necessarily be a
classical Hamiltonian system employing smooth structures. [To find
counter examples is a problem worth giving serious consideration.]

The present work has brought us on the threshold of a formalism that
provides for an autonomous development of QM as a universal
mechanics. Development of such a formalism will be the burden of
paper II.

\vspace{.2in} \noindent \textbf{Acknowledgements}

\vspace{.1in} The author thanks M. Dubois-Violette and M.J.W. Hall
for their critical comments on the paper ``Noncommutative geometry
and unified formalism for classical and quantum mechanics" (I.I.T.
Kanpur preprint, 1993) and (Dass [25]) respectively and K.R.
Parthasarathy, R.Sridharan and V. Balaji for helpful discussions.

\vspace{.2in} \noindent \textbf{\large Appendix : Study of
Implications of the Equations (85) and (86)}

\vspace{.12in} Let $V^{(i)} = \mathcal{L}(\sca^{(i)})$ (i= 1,2) be
the space of linear mappings of the superalgebra $\sca^{(i)}$ into
itself and $ V = V^{(1)} \otimes V^{(2)}$. The equations (85) and
(86) are equations in the space V. Eq.(85) is of the form
\begin{eqnarray} x(A,C) \otimes v(B) = y(A,C) \otimes w(B)
\end{eqnarray}
where A,C are arbitrary homogeneous elements in $\sca^{(1)}$ and B
is a general element of $\sca^{(2)}$. The objects x( . , . ) and
y( . , . ) are bilinear and skew-symmetric mappings  $\sca^{(1)}
\times \sca^{(1)} \rightarrow V^{(1)}$ and  the objects  v(.)
and w(.) are linear mappings $ \sca^{(2)} \rightarrow V^{(2)}$.

The most general implication of Eq.(102) can be expressed in the
form of the relations
\begin{eqnarray} a_1 x = a_2 y; \ \ a_3 v = a_4 w \end{eqnarray}
where the parameters $a_i \in \mathbb{C}$ (i = 1,..,4) satisfy the
condition \begin{eqnarray} a_1 a_3 = a_2 a_4. \end{eqnarray}

Now, the quantities $x(A,C) = \mu_1 ( \{ A,C \}_1)$ and $ w(B) =
Y_B^{(2)}$ cannot be identically zero [this follows from the
non-degeneracy of the symplectic forms of the superalgebras
$\sca^{(i)}$ (i=1,2)]. It follows that, in Eq.(103), we must have
$a_2 \neq 0, \ a_3 \neq 0$ We now have, from Eq.(104)
\[ \frac{a_1}{a_2} = \frac{a_4}{a_3} = a, \ \textnormal{say} \]
and, from Eq.(103)
\begin{eqnarray} y = ax, \ \ v = a w. \end{eqnarray}
Since the possibilities $ y \equiv 0$ and $ v \equiv 0$ cannot be
ruled out, we must allow the value a= 0.

The equations (105) are precisely the equations (89) and (90) with $
a = \lambda_2$. Similar arguments with Eq.(86) lead to the equations
(87) and (88).

\vspace{.2in} \noindent \textbf{References} {\footnotesize
\begin{description}
\item[[1]] R. Abraham, J.E.  Marsden  (with the assistance of T. Ratiu
and R. Cushman ), Foundations of Mechanics, second ed., Benjamin,
Massachusetts, 1978.
\item[[2]] V. Aldaya, J.A. Azcarraga, Geometrical formulation of
classical mechanics and field theory, Nuov. Cim. \textbf{3} (1980), 1-66.
\item[[3]] L.M. Alonso, Group theoretical foundations of classical and
quantum mechanics, II. Elementary systems, J. Math. Phys.
\textbf{20} (1979), 219-230.
\item[[4]] H. Araki,  Mathematical Theory of Quantized Fields,
Oxford Univ. Press, 1999.
\item[[5]] V.I. Arnold, Mathematical Methods of Classical
Mechanics, Springer-Verlag, New York, 1978.
\item[[6]] A. Ashtekar, J. Lewandowski, Background independent quantum
gravity: a status report, Class. Quant. Grav. \textbf{21} (2004),
R53-R152; arXiv: gr-qc/0404018.
\item[[7]] J.C. Baez, I.E. Segal, Z. Zhou, Introduction to Algebraic
and Constructive Quantum Field Theory,  Princeton University Press,
1992.
\item[[8]] F.A. Berezin, M.S. Marinov, Particle spin dynamics as the
Grassmann variant of classical mechanics, Ann. Phys. \textbf{104}
(1977), 336-362.
\item[[9]] R. Berndt, An Introduction to Symplectic Geometry,
American Mathematical Society, Providence, 2001.
\item[[10]] H.J. Borchers, On revolutionalizing quantum field theory
with Tomita's modular theory, J. Math. Phys. \textbf{41} (2000),
3604.
\item[[11]] M. Born, W. Heisenberg, P. Jordan, Zur quantenmechanik
II, Zs. f. Phys. \textbf{35} (1926), 557.
\item[[12]] M. Born, P. Jordan, Zur quantenmechanik, Zs. f.
Phys. \textbf{34} (1925), 858.
\item[[13]] O. Bratteli, D.W. Robinson, : Operator Algebras and
Quantum Statistical Mechanics I. $C^*-$ and $W^*$-algebras, Symmetry
Groups, Decomposition of States. Springer, New York,1979; II.
Equilibrium States; Models in Quantum Statistical Mechanics, ibid
1981.
\item[[14]] J.F. Cari$\tilde{n}$ena, M. Santander, On the projective unitary
representations of connected Lie groups, J. Math. Phys. \textbf{16}
(1975), 1416.
\item[[15]] J. Caro, L.L. Salcedo,  Impediments to mixing
classical and quantum dynamics, Phys. Rev. \textbf{A60} (1999), 842;
arXiv : quant-ph/9812046 v2.
\item[[16]] H. Cartan, S. Eilenberg, Homological Algebra,
Princeton University Press, 1956.
\item[[17]] A. Chemseddine, A. Connes,  Universal formula for noncommutative
geometry actions: Unification of gravity and the standard model,
Phys. Rev. Letters \textbf{77} (1996), 486804871.
\item[[18]] A. Chemseddine, A. Connes, The spectral action principle, Comm.
Math. Phys. \textbf{186} (1997), 731-750.
\item[[19]] A. Connes, Noncommutative  Geometry, Academic Press, New
York, 1994.
\item[[20]] A. Connes, Noncommutative  geometry and reality, Jour. Math.
Phys. \textbf{36} (1995), 6194-6231.
\item[[21]] A. Connes, Gravity coupled matter and the foundations of
noncommutative geometry, Comm. Math. Phys. \textbf{182} (1995),
155-176.
\item[[22]] A. Connes, M. Marcolli , Noncommutative Geometry,
Quantum Fields and Motives, Hindustan Book Agency, New Delhi, 2008.
\item[[23]] T. Dass, Symmetries, Gauge Fields, Strings and
Fundamental Interactions, vol. I: Mathematical Techniques in Gauge
and String Theories, Wiley Eastern Limited, New Delhi, 1993.
\item[[24]] T. Dass, Towards an autonomous formalism for quantum
mechanics, arXiv : quant-ph/0207104 (2002).
\item[[25]] T. Dass, Consistent quantum-classical interaction
and solution of the measurement problem, arXiv : quant-ph/0612224
(2006).
\item[[26]] T. Dass, Universality of quantum symplectic structure,
arXiv : 0709.4312 (math.SG) (2007).
\item[[27]] T. Dass, Supmech : the geometro-statistical formalism
underlying quantum mechanics, arXiv : 0807.3604 v3 (quant-ph), 2008.
\item[[28]] T. Dass, Supmech : a noncommutative geometry based
universal mechanics accommodating classical and quantum
mechanics, in Proceedings of the International Congress of
Mathematicians, Hyderabad (2010), Abstracts, p. 387.
\item[[29]] A. Dimakis, F. M$\ddot{u}$ller-Hoissen, Quantum
mechanics as noncommutative symplectic geometry, J. Phys. A : Math.
Gen. \textbf{25} (1992) 5625-5648.
\item[[30]] P.A.M. Dirac, The fundamental equations of quantum mechanics.
Proc. Roy. Soc. \textbf{A 109} (1926), 642.
\item[[31]] A.E.F. Djemai, Introduction to Dubois-Violette's noncommutative
differential geometry, Int. J. Theor. Phys. \textbf{34} (1995), 801.
\item[[32]] J.F. Donoghue, E. Golowich E., B.R. Holstein, Dynamics of
the Standard Model, Cambridge University Press, 1994.
\item[[33]] D.A. Dubin, M.A. Hennings, Quantum Mechanics, Algebras
and Distributions, Longman Scientific and Technical, Harlow, 1990.
\item[[34]] M. Dubois-Violette, Noncommutative differential geometry,
quantum mechanics and gauge theory, in Lecture Notes in Physics
\textbf{375},  Springer, Berlin, 1991.
\item[[35]] M. Dubois-Violette, Some aspects of noncommutative
differential geometry, in New Trends in Geometrical and Topological
Methods (Madeira 1995) pp 145-157. Cont. Math. textbf{203}, Amer.
Math. Soc., 1997; arXiv: q-alg/9511027.
\item[[36]] M. Dubois-Violette, Lectures on graded differential algebras and
noncommutative geometry, in Noncommutative Differential Geometry and
its Application to Physics (Shonan, Japan, 1999), pp 245-306. Kluwer
Academic Publishers, 2001; arXiv: math.QA/9912017.
\item[[37]] M. Dubois-Violette, R. Kerner, J. Madore, Noncommutative
differential geometry of matrix algebras. \emph{J. Math. Phys.}
\textbf{31} (1990), 316.
\item[[38]] M. Dubois-Violette, R. Kerner, J. Madore, Supermatrix
geometry, Class. Quant. Grav. \textbf{8} (1991), 1077.
\item[[39]] M. Dubois-Violette, A. Kriegl, Y. Maeda, P.W. Michor,
Smooth *-algebras, Progress of Theoretical Physics Suppl.
\textbf{144} (2001), 54-78; arXiv : math.QA/0106150.
\item[[40]] G.E. Emch, Algebraic Methods in Statistical
Mechanics and Quantum Field Theory, Wiley, New York, 1972.
\item[[41]] G.E. Emch, Mathematical and Conceptual Foundations of
20th Century Physics, North-Holland, Amsterdam, 1984.
\item[[42]] G. Giachetta, L. Mangiarotti, G. Sardanshvily,  Geometric
and Algebraic Topological Methods in Quantum Mechanics, World
Scientific, Singapore, 2005.
\item[[43]] J. Glimm, A. Jaffe, Quantum Physics: a Functional
Integral Point of View, Springer Verlag, New York, 1981.
\item[[44]] J.M. Gracia Bondia, J.C. Varilly, H. Figuerra, Elements
of Noncommutative Geometry,  Birkha$\ddot{u}$ser, Boston, 2001.
\item[[45]] M.B.Green, J.H. Schwarz, E. Witten, Superstring Theory,
Vol. I,II, Cambridge University Press, 1987.
\item[[46]] W. Greub, Linear Algebra, Springer Verlag, New
York, 1975.
\item[[47]] W. Greub, Multilinear Algebra,
Springer-Verlag, New York, 1978.
\item[[48]] H. Grosse, G. Reiter, Graded differential geometry of
graded matrix algebras.  arXiv: math-ph/9905018 (1999).
\item[[49]] V. Guillemin, S. Sternberg, Symplectic Techniques in
Physics,  Cambridge University Press, 1984.
\item[[50]] R. Haag, Local Quantum Physics, Springer, Berlin, 1992.
\item[[51]] R. Haag, D. Kastler,  An algebraic approach to quantum field
theory, J. Math. Phys. \textbf{5} (1964), 848.
\item[[52]] W. Heisenberg, $\ddot{U}$ber quantentheoretische umdeutung
kinematicher und mechanischer beziehungen (Quantum-theoretical
re-interpretation of kinematic and mechanical relations). Zs. f.
Phys. \textbf{33} (1925), 879-893.
\item[[53]]  A. Ya. Helemskii, The Homology of Banach and Topological
Algebras, Kluwer Academic Publishers, 1989.
\item[[54]] D. Hilbert, Mathematical problems, lectures delivered before
the International Congress of Mathematicians in Paris in 1900,
translated by M.W. Newson. \emph{Bull. Amer. Math. Soc.} \textbf{8}
(1902), 437.
\item[[55]] S.S. Horuzhy, Introduction to Algebraic Quantum Field
Theory,  Kluwer Academic Publishers, Dordrecht, 1990.
\item[[56]] S. Iguri, M. Castagnino, The formulation of quantum mechanics
in terms of nuclear algebras, Int. J. Theor. Phys. \textbf{38}
(1999), 143.
\item[[57]] P. Jordan, J. von Neumann, E. Wigner,  On an Algebraic
Generalization of the Quantum Mechanical Formalism. \emph{Ann.
Math.} \textbf{35} (1934), 29.
\item[[58]] R. Kerner, Graded noncommutative geometries,
J. Geom. Phys, \textbf{11} (1993), 325.
\item[[59]] G. Landi, An Introduction to Noncommutative Spaces and
Their Geometries, Springer, Berlin, 1997.
\item[[60]] J. Madore, An Introduction to Noncommutative Geometry and
its Applications, Cambridge University Press, 1995.
\item[[61]] Y. Matsushima, Differentiable Manifolds, Marcel Dekker,
New York, 1972.
\item[[62]] P. -A. Meyer, Quantum Probability for Probabilists,  second
ed., Springer-Verlag, Berlin, 1995.
\item[[63]] L. Pittner, Algebraic Foundations of
Non-Commutative Differential Geometry and Quantum Groups,
Springer-Verlag, Berlin, 1996.
\item[[64]] J. Polchinski, String Theory, Vol.I,II, Cambridge University
Press, 1998.
\item[[65]] C. Rovelli, Quantum Gravity,
Cambridge University Press, 2004.
\item[[66]] L.L. Salcedo, Absence of classical and quantum mixing,
\emph{Phys. Rev.} \textbf{A54} (1996), 3657-3660 [ArXiv :
hep-th/9509089].
\item[[67]] M. Scheunert, The Theory of Lie Superalgebras, Lecture
Notes in Mathematics, \textbf{716}, Springer-Verlag, Berlin, 1979.
\item[[68]] M. Scheunert, (i) Generalized Lie algebras,
J. Math. Phys. \textbf{20} (1979), 712-720; (ii) Graded tensor
calculus. J. Math. Phys.  \textbf{24} (1983), 2658-2670.
\item[[69]] M. Scheunert, R.B. Zhang,  Cohomology of Lie superalgebras
and of their generalizations. J. Math. Phys. \textbf{39} (1998),
5024-5061 [arxiv : q-alg/9701037].
\item[[70]] I.E. Segal, Postulates for general quantum mechanics,
Ann. of Math. \textbf{48} (1947), 930.
\item[[71]] I.E. Segal, Mathematical Problems of Relativistic
Physics, (with an appendix by G.W. Mackey). Amer. Math. Soc.,
Providence, 1963.
\item[[72]] J.-M. Souriau, Structure of Dynamical Systems:
a Symplectic View of Physics, Birkh$\ddot{a}$user, Boston, 1997.
\item[[73]] E.C.G.  Sudarshan, N. Mukunda, Classical Dynamics : A
Modern Perspective, Wiley, New York, 1974.
\item[[74]] F. Treves, Topological Vector Spaces, Distributions and
Kernels, Academic Press, San Diego, 1967.
\item[[75]] C.A. Weibel, An Introduction
to Homological Algebra, Cambridge University Press, 1994.
\item[[76]] S. Weinbeg, The Quantum Theory of fields, vol.
I,II,III, Cambridge Univ. Press, 2000.
\item[[77]] A.S. Wightman, Hilbert's sixth problem: Mathematical
treatment of the axioms of physics, Proceedings of Symposia in Pure
Mathematics \textbf{28} (1976), 147.
\item[[78]] N. Woodhouse, Geometric Quantization, Clarendon Press,
Oxford, 1974.
\item[[79]] K. Yosida, Functional Analysis, Narosa
Publishing House, New Delhi, 1979.
\end{description}
\end{document}